\documentclass[twocolumn]{aastex631}

\usepackage{savesym}
\usepackage{graphicx}	% Including figure files
\usepackage{amsmath}	% Advanced maths commands
\usepackage{amssymb}	% Extra maths symbols

\renewcommand{\figurename}{Fig.}

\savesymbol{longtable}
\usepackage{xcolor}

\usepackage[utf8]{inputenc}
\usepackage[english]{babel}

\restoresymbol{SUB}{longtable}

\usepackage{physics}

\begin{document}

\title{Nonlinear feedback of the electrostatic instability on the blazar-induced pair beam and GeV cascade}

\author[0009-0008-0835-2795]{Mahmoud Alawashra}
\affiliation{Institute for Physics and Astronomy, University of Potsdam, D-14476 Potsdam, Germany}
\email{mahmoud.s.a.alawashra@uni-potsdam.de}

\author[0000-0001-7861-1707]{Martin Pohl}
\affiliation{Institute for Physics and Astronomy, University of Potsdam, D-14476 Potsdam, Germany}
\affiliation{Deutsches Elektronen-Synchrotron DESY, Platanenallee 6, 15738 Zeuthen, Germany}
\email{martin.pohl@desy.de}

\received{18-Oct-2023}
\revised{26-Jan-2024}
\accepted{30-Jan-2024}
\published{-}
\submitjournal{The Astrophysical Journal}

\begin{abstract}

Relativistic pair beams produced in the cosmic voids by TeV gamma rays from blazars are expected to produce a detectable GeV-scale cascade that is missing in the observations. The suppression of this secondary cascade implies either the deflection of the pair beam by intergalactic magnetic fields or, alternatively, an energy loss of the beam due to the beam-plasma instability. Here, we study how the beam-plasma instability feeds back on the beam, using a realistic two-dimensional beam distribution. We find that the instability broadens the beam opening angles significantly without any significant energy loss, thus confirming a recent feedback study on a simplified one-dimensional beam distribution. However, narrowing diffusion feedback of the beam particles with Lorentz factors less than $10^6$ might become relevant even though initially it is negligible. Finally, when considering the continuous creation of TeV pairs, we find that the beam distribution and the wave spectrum reach a new quasi-steady state, in which the scattering of beam particles persists and the beam opening angle may increase by a factor of hundreds. Understanding the implications on the GeV cascade emission requires accounting for inverse-Compton cooling.

\end{abstract}


\keywords{gamma rays: general -- instabilities -- Blazar -- relativistic processes -- waves}

%%%%%%%%%%%%%%%%%%%%%%%%%%%%%%%%%%%%%%%%%%%%%%%%%%

%%%%%%%%%%%%%%%%% BODY OF PAPER %%%%%%%%%%%%%%%%%%

\section{Introduction}

Blazars are active galactic nuclei with their relativistic jet pointing toward Earth. Observations of the Fermi-LAT telescope and the imaging atmospheric Cerenkov telescopes (such as VERITAS, MAGIC, and HESS) show bright GeV-TeV $\gamma$-ray emission from several blazars. During their propagation through the intergalactic medium (IGM), those very high energy $\gamma$-rays interact with the extragalactic background light (EBL), producing a focused beam of electron-positron pairs, that are anticipated to dissipate their energies via inverse Compton scattering on the cosmic microwave background (CMB) \citep{1967PhRv..155.1408G,1970RvMP...42..237B}. 

Although primary $\gamma$-rays with energies of a few TeV would initiate an electromagnetic cascade in the GeV energy range, such emissions seem to be absent from the $\gamma$-ray spectra of certain blazars \citep{2009PhRvD..80l3012N} and possibly the isotropic $\gamma$-ray background \citep{2023arXiv230301524B}. One possible explanation for the absence of the GeV cascade emission from the $\gamma$-ray spectra of blazars is deflection of the TeV pairs by the intergalactic magnetic fields (IGMF) \citep{PhysRevD.80.023010,PhysRevD.80.123012,2010Sci...328...73N,2011A&A...529A.144T,Takahashi_2011,Vovk_2012,Durrer_2013}. This deflection results in an extended emission or/and a time delay of the cascade emission. In this case, the observed blazar spectra are used to put lower limits on the strength of the IGMF.

The only alternative solution for the missing GeV cascade emission within the standard model of physics is beam energy loss by collective beam-plasma instabilities that is faster than inverse Compton cooling on the CMB. However, whether the non-linear evolution of those instabilities is efficient in taking a significant fraction of the beam energy is still debated \citep{Broderick_2012,Schlickeiser_2012,Miniati_2013,Schlickeiser_2013,2014ApJ...790..137B,Sironi_2014,Chang_2014,Supsar_2014,2016ApJ...833..118C,2016A&A...585A.132K,2017A&A...607A.112R,Vafin_2018,Vafin_2019,AlvesBatista:2019ipr,shalaby_broderick_chang_pfrommer_puchwein_lamberts_2020,Perry_2021,Alawashra_2022}.

\citet{Vafin_2018} calculated the linear growth rate of the electrostatic instability using a realistic pair distribution generated by the annihilation of high-energy gamma rays with the extragalactic background light. Their results demonstrated that the finite angular spread of the beam plays a decisive role in shaping the unstable electrostatic modes. Specifically, their results revealed that the fastest growth rates occurred for wave vectors that are quasi-parallel to the beam direction, while growth rates at oblique directions are smaller compared to the peak values.

Previous studies of the blazar-induced pair beam instabilities didn't consider the instability feedback on the pair beam particles. \citet{Perry_2021} studied this feedback for the first time in the context of blazar-induced pair beam electrostatic instability. Their findings imply that the back reaction of the electrostatic unstable waves on the pair beam widens the beam opening angles by around one order of magnitude without any significant energy loss.

In this study, unlike the simplified one-dimensional beam distribution used in \citet{Perry_2021}, we use a two-dimensional realistic beam distribution to explore the influence of the instability feedback on the beam. Specifically, we use the beam profile at a distance of 50 Mpc from the blazar found in \citet{Vafin_2018}. This treatment enables us to look at the feedback influence on the pairs that have the relevant Lorentz factors for cascade emission in the GeV band.

The instability feedback is described as Fokker-Planck diffusion both in momentum and angular space. This treatment was simplified in the analysis by \citet{Perry_2021}, by evaluating only the initially dominant angular widening diffusion and neglecting the other effects involving the momentum diffusion and angular narrowing ($D_{\theta p}$). Here, we check rigorously this assumption by using the 2D spectrum of the expanded beam under the dominant feedback to analyse the possible impact of the momentum diffusion on the beam energy and whether the beam narrowing is still negligible.

The blazar-induced pair beam-plasma instability significantly outpaces other factors that could change the beam profile, such as inverse Compton cooling and pair production. Whereas previous works have predominantly focused on assessing the instability's impact on a stationary beam profile, we incorporate the continuous production of TeV pairs into the transport equation of the beam, in addition to the diffusion terms. 

The structure of this paper is as follows. In section \ref{sec:2}, we introduce the pair beam realistic 2D profile that we used in this study. In section \ref{sec:3}, we present the quasilinear theory of the beam-plasma system, we introduce the linear growth rate of the electrostatic instability, the time evolution of the electrostatic waves spectrum and the Fokker-Planck diffusion of the beam distribution. Finally, we demonstrate the numerical simulation and the results in section \ref{sec:4} and conclude in section \ref{sec:con}.

\section{Blazar-induced pair beam distribution}\label{sec:2}

The pair-beam distribution function is the crucial quantity that determines the beam-plasma instability growth rate \citep{Vafin_2018}. Thus, using the realistic spectrum of blazar-induced pair beams is essential for examining the influence of the beam-plasma instability on the beam and the GeV-scale cascade emission. In this study, we used the realistic beam distribution at a distance of 50 Mpc from the blazar, as reported in \citet{Vafin_2018}. Here, we introduce this beam spectrum and explain the ingredients used to find it.

The propagation of the beam distribution in the IGM is driven by two primary factors. The first one is the pair's production due to the interaction of the high-energy gamma rays with the EBL, along with their subsequent cooling processes. The second effect is the dispersion of the primary gamma-ray flux with the propagation distance, leading to an inverse proportionality of the beam density with the square of the distance from the blazar. These two fundamental mechanisms collectively shape the evolution of the pair-beam distribution along the propagation distance in the IGM.

The two effects have been combined in \citet{Vafin_2018}, neglecting the IC cooling, to calculate the accumulated pair spectrum over the IC cooling mean-free path of pairs with Lorentz factor of $10^7$ starting at the distance 50 Mpc from the blazar. The neglect of IC cooling is driven by the necessity to investigate beam-plasma instabilities that provide the dominant energy loss. They used an intrinsic power-law gamma-ray spectrum with a spectral index of 1.8 and a cut-off step function at the energy of 50 TeV.

We define the normalized beam momentum distribution
\begin{equation}\label{eq:norm}
    \int d^3p \text{ } f(p,\theta) = n_b,
\end{equation}
where $n_b$ is the pair-beam density. The density at 50 Mpc from the blazar is estimated as $n_b = 3\times 10^{-22}$ cm$^{-3}$ \citep{Vafin_2018}. Factorizing the distribution as
\begin{equation}\label{eq:1}
    f(p,\theta) = \frac{d^3f}{dp^3} = \frac{1}{ 2 \pi m_ec\beta p^2}f_\gamma(\gamma)f_{\cos{\theta}}(\gamma,\theta),
\end{equation}
where $f_{\cos{\theta}}(\gamma,\theta)$ is an angular differential part, $f_\gamma(\gamma)$ is a momentum differential part, and $\beta$ is the normalized speed. The angular part is approximated by a Gaussian \citep{Broderick_2012,Miniati_2013,Vafin_2018}
\begin{equation}\label{eq:2}
    f_{\cos{\theta}}(\gamma,\theta) =\frac{2}{ \Delta\theta^2}\exp{-\big(\frac{\theta}{\Delta\theta}\big)^2},
\end{equation}
with the angular spread of $\Delta\theta = \frac{1}{\gamma}$. The momentum part, $f_\gamma(\gamma)$, is given by eq.\ref{eq:fbg} in the appendix \ref{app:fbg}, where we replaced the sharp step-function cut-off used in \citet{Vafin_2018} by an exponential cut-off at the Lorentz factors higher than $6\times 10^6$ using a part of a logarithmic Gaussian as shown in Fig.\ref{fig:fbg}.

\begin{figure}
\centering
    \includegraphics[width=\columnwidth]{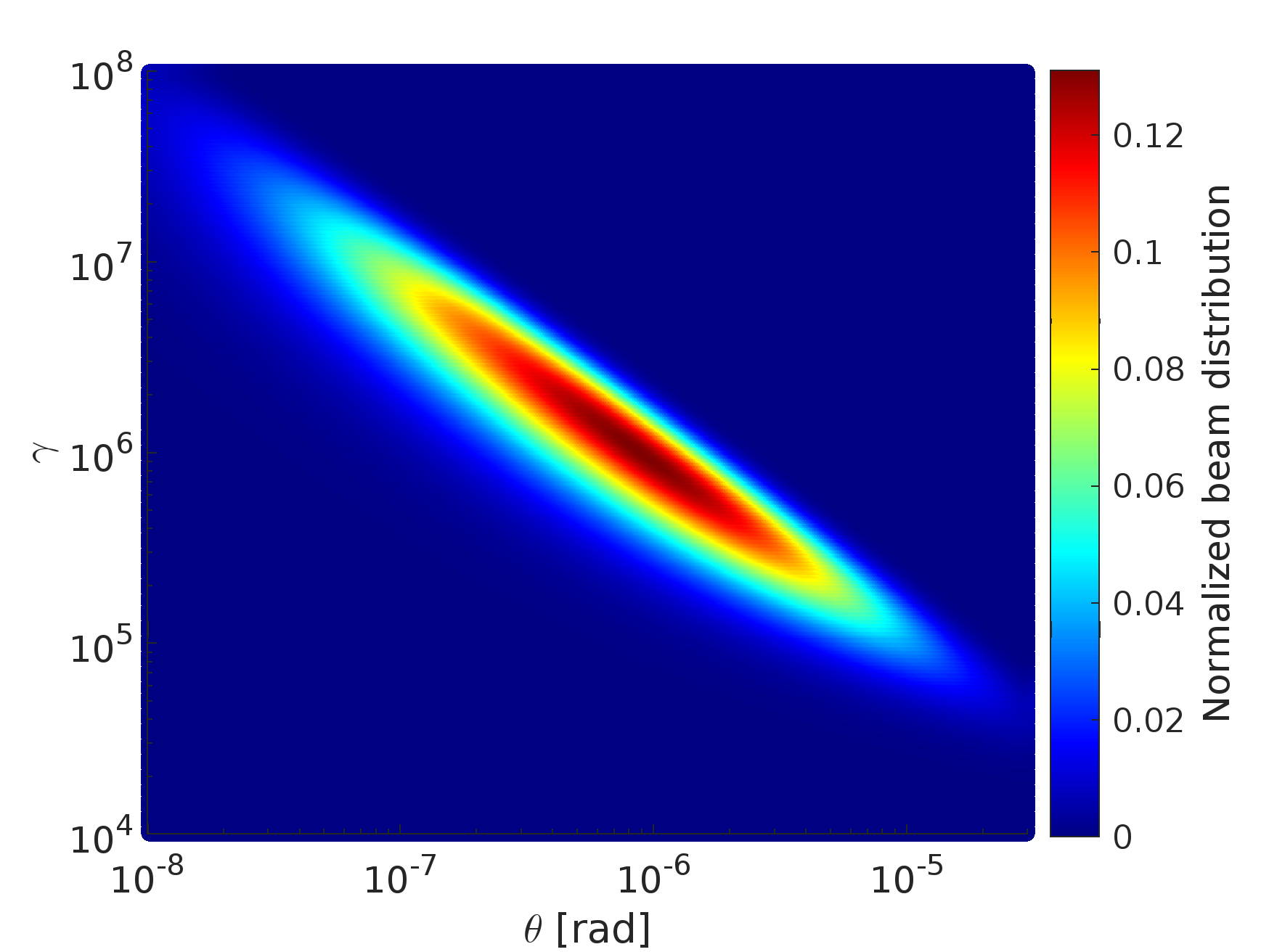}
    \caption{The normalized initial beam distribution $2\pi p^3\theta^2 \,f(p,\theta)/n_b$ at distance 50 Mpc from the blazar \citep{Vafin_2018}.}
    \label{fig:f0}
\end{figure}

The initial normalized realistic beam spectrum is shown in Fig.\ref{fig:f0}, and the main energy bulk of the pair is located at Lorentz factors of a few $10^{6}$. We also see that the pairs are concentrated in a narrow band around the production angles of $\theta \sim \gamma^{-1}$. We will use this distribution to find the linear growth rate of the instability in section \ref{sec:3.1} and as the initial condition for the Fokker-Planck simulation of the instability feedback in section \ref{sec:4}.

\section{Quasilinear theory of the beam-plasma system}\label{sec:3}

The beam-plasma instabilities manifest in both electrostatic and electromagnetic modes, including the two-stream instability ($\vb{k}\cross{\delta\vb{E}}=0$ where $\delta\vb{E}$ is the perturbed electric field), the transverse Weibel, and filamentation modes ($\vb{k}\cdot{\delta\vb{E}}=0$) \citep{doi:10.1063/1.3514586}. The electrostatic modes dominate the wave spectrum for the blazar-induced TeV beams, whereas Weibel-type modes are suppressed \citep{2005PhRvE..72a6403B,2017A&A...607A.112R}. Consequently, we consider only the electrostatic oblique modes, which is sufficient to recover the essential physics \citep{2016ApJ...833..118C}.

In section \ref{sec:3.1}, we present the linear growth rate of the electrostatic instability. In section \ref{sec:3.2}, we introduce the balance equation for waves. Lastly, in section \ref{sec:3.3}, we introduce the Fokker-Planck diffusion equation describing the instability feedback on the beam distribution.

\subsection{Electrostatic linear growth rate}\label{sec:3.1}

In the kinetic regime that is applicable for blazar-induced pair beams \citep{Miniati_2013}, the linear growth rate of an unstable wave with a wave vector $\boldsymbol{k}$ can be found by \citep{1990}
\begin{equation}
    \omega_i(\boldsymbol{k}) = \omega_p \frac{2\pi^2 e^2}{k^2} \int d^3\boldsymbol{p} \left(\boldsymbol{k}\cdot\frac{\partial f(\boldsymbol{p})}{\partial \boldsymbol{p}}\right) \delta(\omega_p - \boldsymbol{k}\cdot\boldsymbol{v}_b),
\end{equation}
where $\omega_p = (4\pi n_e e^2/m_e)^{1/2}$ is the plasma frequency of the intergalactic background plasma with density $n_e$. Here, we neglected the influence of intergalactic medium temperature, $T_e$, on the plasma frequency, $\omega$, expressed as $\omega^2 = \omega_p^2 + 3 k^2 v_\text{th,e}^2 \approx \omega_p^2$, where $v_\text{th,e}$ is the IGM electros' thermal speed. For an intergalactic temperature of $10^4$ K, the plasma frequency experiences a negligible shift of around $2.6 \times 10^{-6}$ ($\omega = \omega_p (1+2.6 \times 10^{-6})$), insignificantly affecting linear growth rate calculations. However, it's important to mention that intergalactic medium temperature may significantly impact the non-linear evolution of waves, as discussed in studies of non-linear landau damping by \citep{Miniati_2013,Chang_2014,Vafin_2019}.

Exploiting the cylindrical symmetry around the beam propagation axis ($z$-axis in our case), we fixed the wave vector of the electrostatic waves to $\boldsymbol{k} = (k_\perp,0,k_{||})$, where $k_\perp$ and $k_{||}$ are the perpendicular and the parallel components to the beam propagation direction respectively. After integrating over the azimuthal angle of the beam we get the following
\begin{equation}\label{eq:growth}
\begin{split}
    \omega_i(k_\perp,k_{||}&)  =  \pi \frac{\omega_p}{n_e} \left(\frac{\omega_p}{kc}\right)^3 \int dp m_e c \text{ } p \int_{\theta_1}^{\theta_2}d\theta \\ & \times  \frac{-2f(p,\theta)\sin{\theta+(\cos{\theta}-\frac{kv_b}{\omega_p}\cos{\theta'}})\frac{\partial f(p,\theta)}{\partial \theta}}{[(\cos{\theta_1}-\cos{\theta})(\cos{\theta}-\cos{\theta_2})]^{1/2}},
\end{split}
\end{equation}
where the boundaries are
\begin{equation}\label{eq:growthlimit}
    \cos{\theta_{1,2}} = \frac{\omega_p}{kv_b} \left( \cos{\theta'}\pm \sin{\theta'} \sqrt{ \left(\frac{kv_b}{\omega_p}\right)^2-1}\right).
\end{equation}
Here $\theta'$ is the wave vector angle with the beam propagation direction ($z$-axis), $\theta$ is the angle between momentum and the beam axis, and $v_b \simeq c (1-\frac{1}{2\gamma^2})$ is the particle speed.

\begin{figure}
\centering
    \includegraphics[width=\columnwidth]{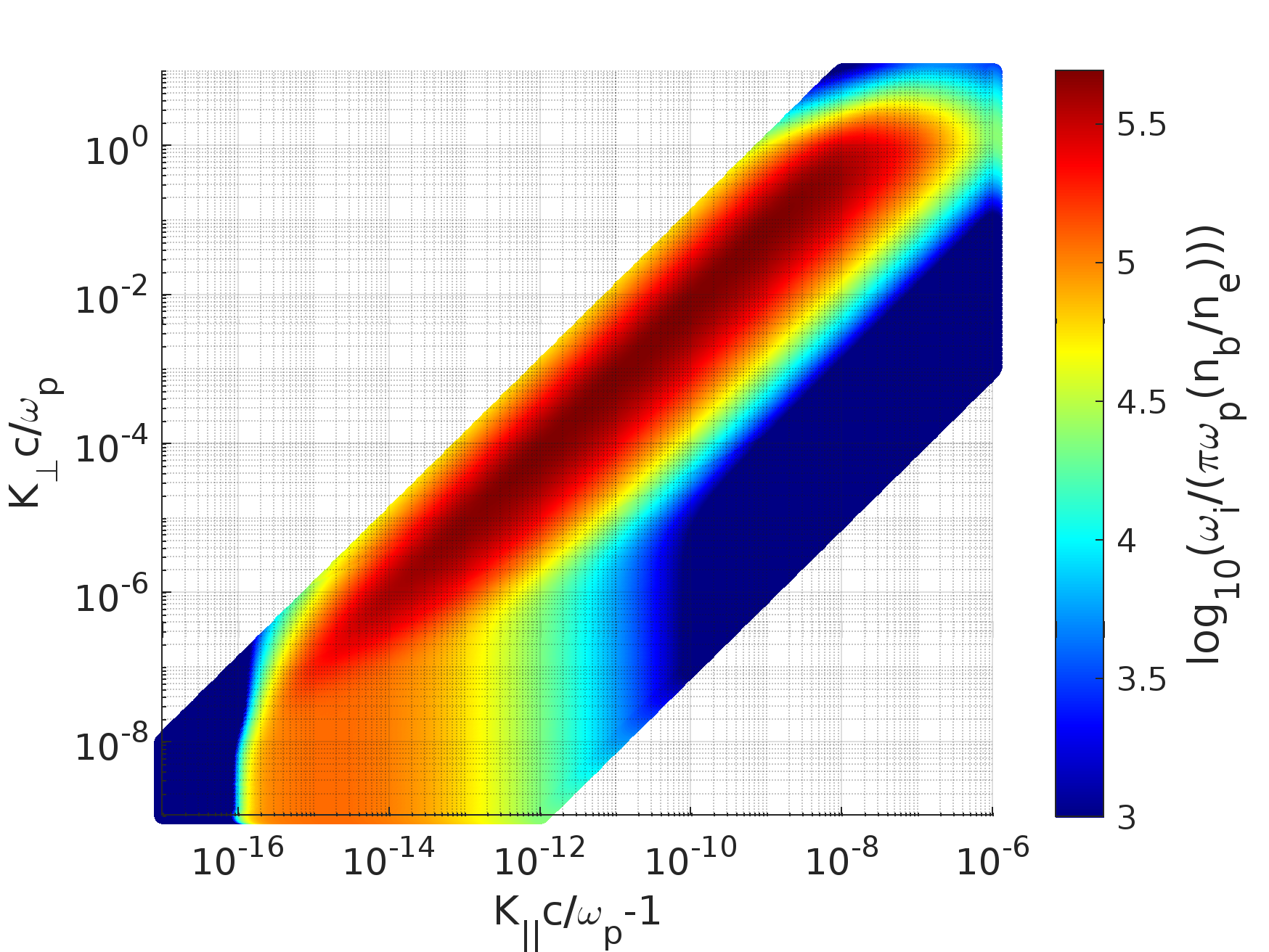}
    \caption{Normalized linear growth rate using the realistic beam distribution function. White areas denote stable modes.}
    \label{fig:wi}
\end{figure}

In Fig.\ref{fig:wi}, we present the linear growth rate (eq.\ref{eq:growth} - \ref{eq:growthlimit}), using the beam distribution introduced in section \ref{sec:2}, the density of IGM electron as $n_{e}$ = $10^{-7} (1+z)^3$cm$^{-3}$, and a redshift $z=0.15$. Previous treatments in the literature used the approximation ($v_b = c$) when evaluating eq.\ref{eq:growth} \citep{Miniati_2013,Vafin_2018,Perry_2021}. However, we found that in the regimes of wave numbers with $\left(ck_\perp/\omega_p\right)^2 \sim \left(\frac{ck_{||}}{\omega_p}-1\right)$ and $\left(ck_\perp/\omega_p\right)^2 << \left(\frac{ck_{||}}{\omega_p}-1\right)$, the difference between the particle speed and the speed of light becomes relevant. We have taken this difference into account in our calculations of Fig.\ref{fig:wi}.

In Fig.\ref{fig:wi}, we see that the growth rate is maximal and constant in the range of perpendicular wave numbers, $10^{-6} < ck_\perp/\omega_p<1$, with a sharp drop at the oblique angles $ck_\perp/\omega_p \sim 1$, whereas for the parallel modes, $k_\perp c/\omega_p<10^{-6}$, it is smaller by around a factor of 3. Note that the growth rate of the parallel modes is sensitive to the beam distribution, for a simplified monoenergetic Gaussian the growth of the parallel modes is larger than the quasi-parallel ones \citep{Perry_2021}, where it's smaller for the distribution we used and for a Maxwell–Jüttner distribution \citep{2016ApJ...833..118C}.

The turnover of the linear growth rate spectral shape around the wave numbers of $\left(ck_\perp/\omega_p\right)^2 \sim \left(\frac{ck_{||}}{\omega_p}-1\right)$ is due to the change in the corresponding resonant beam angles. In the regime of $\left(ck_\perp/\omega_p\right)^2 >> \left(\frac{ck_{||}}{\omega_p}-1\right)$, the resonant beam angles are constrained by the minimum angles of $\theta_1 = (ck_{||}/\omega_p-1)/(ck_\perp/\omega_p)$ and large $\theta_2$. In the regime of $\left(ck_\perp/\omega_p\right)^2 << \left(\frac{ck_{||}}{\omega_p}-1\right)$, the resonant angles boundaries approach each other to a $ck_\perp/\omega_p$ independent value of $\theta_{1,2} \sim \sqrt{2}\left[\left(\frac{ck_{||}}{\omega_p}-1\right) - \frac{1}{2\gamma^2}\right]^{1/2}$.

The maximum growth rate, $\omega_{i,\text{max}}  \sim 6.7 \times 10^{-8} \ \text{s}^{-1}$, is much faster than the IC cooling rate of the beam, $\tau^{-1}_\text{IC}(\gamma) \approx \gamma \times 1.3 \times 10^{-20} (1+z)^4 \ \text{s}^{-1}$.  However, the instability-induced energy-loss rate significantly depends on the nonlinear evolution of the instability \citep{Miniati_2013,Schlickeiser_2013,Chang_2014,Vafin_2019}. 

In this study we focus on the instability feedback, therefore we will consider only the linear regime of the instability and neglect the restrictions on the growth of the waves due to non-linear interactions. In the next section, we briefly introduce the linear evolution equation of unstable waves and levels of the unstable wave's energy density where the nonlinear processes become relevant.

\subsection{Evolution of the wave spectrum}\label{sec:3.2}

The quasi-linear evolution of the wave spectrum for homogeneous plasma is governed by the following equation 
\begin{equation}\label{eq:W}
    \frac{\partial W (\vb{k})}{\partial t} = 2 (\omega_i (\vb{k}) + \omega_c(k)) W(\vb{k}), 
\end{equation}
where $W(\vb{k})$ is the spectral energy density of the electric field oscillations, $\omega_i(\vb{k})$ is the linear growth rate as defined in section \ref{sec:3.1}, and $\omega_c$ is the collisional damping rate \citep{Tigik_2019},
\begin{equation}\label{eq:damp}
     \omega_c(k) = - \omega_p \frac{g}{6\pi^{3/2}}\frac{1}{(1+3k^2\lambda_D^2)^3}.
\end{equation}
Here $g=(n_e\lambda_D^3)^{-1}$ is the plasma parameter, $\lambda_D = 6.9 \text{ cm} \sqrt{\frac{T_e/ K}{n_e/\text{cm}^{-3}}} $ is the Debye length, $n_{e} = 10^{-7} (1+z)^3 \text{cm}^{-3}$ is the density of IGM electrons, and $T_e = 10^{4} K$ is their temperature. We start integrating eq.(\ref{eq:W}) at the very low thermal fluctuations level.

The collisional damping rate given by eq.\ref{eq:damp} is approximately 20 times smaller than the approximation employed in other studies (i.e. \citet{Miniati_2013,Vafin_2019,Perry_2021}), which did not account for the microscopic wave-particle interactions. Those interactions were included under the generalized weak turbulence theory in \citet{PhysRevE.93.033203}, deriving an accurate general kinetic formulation of the collisional damping rate of the electrostatic plasma waves, that was used in \citet{Tigik_2019} to find the collisional damping rate (eq.\ref{eq:damp}). 

The total electric field energy density is calculated by
\begin{equation}\label{eq:Wtot}
    W_{\text{tot}} = 2 \pi \int dk_\perp k_\perp \int dk_{||} W(k_\perp,k_{||}).
\end{equation}
Accounting for the energy equipartition between kinetic electrostatic fluctuations, the energy loss rate of the beam due to the growth of the electrostatic waves at time $t$ is given by \citep{Vafin_2018}
\begin{equation}
\begin{split}
    \frac{dU_b}{dt}(t) = & - 2 \frac{dW_{\text{tot}}}{dt}(t) \\ = & -8\pi\int dk_\perp k_\perp \int dk_{||} W(k_\perp,k_{||},t) \omega_{i}(k_\perp,k_{||},t),
\end{split}
\label{eq:lossrate}
\end{equation}
where the total beam energy density is defined as
\begin{equation}
    U_b = 2 \pi \int dp p^2 \int d\theta \sin{\theta} m_e c^2 \gamma f(p,\theta).
\end{equation}

In the previous section, we found that the modes with the maximum growth, $10^{-6} < ck_\perp/\omega_p<1$, grow at the same rate (Fig.\ref{fig:wi}), maintaining a similar spectral amplitude. However, the energy density in those modes is proportional to their wave number volume element, $2 \pi k_\perp \Delta k_\perp \Delta k_{||}$ (eq.\ref{eq:Wtot}). Therefore, we can focus on the quasi-parallel and oblique modes, $10^{-3} < ck_\perp/\omega_p<1$, since they dominate the energy density of the unstable mode spectrum. We can also neglect inhomogeneity of the background plasma since it is relevant only for the strictly parallel modes \citep{Perry_2021,shalaby_broderick_chang_pfrommer_puchwein_lamberts_2020}. 

The amplitude of the unstable modes grows exponentially until their wave intensity is high enough to trigger nonlinear processes. One of the main non-linear processes is the modulation instability that moves wave energy from resonant to non-resonant modes. This process operates when the total electric field energy density hits the threshold of $(k\lambda_D)^2n_eT_e$ \citep{Miniati_2013}, resulting in the saturation of the resonant unstable mode at around $10^{-3}$ of the total beam energy we consider here.

Another non-linear process is non-linear Landau damping, where the non-linear scattering of the unstable waves on the background plasma ions results in severe damping of the resonant modes. Non-linear Landau damping becomes effective when the total electric field energy density reaches around $10^{-2}$ of that of the beam \citep{Chang_2014,Vafin_2019}. 

The impact of these non-linear interactions is still uncertain \citep{Schlickeiser_2012,Miniati_2013,Chang_2014,Vafin_2019}. The numerical noise in simulations (such as PIC) is too high, and the numerical growth rate is too small, for a reliable assessment, on account of the very small beam density. Upscaling of the beam density and downscaling the beam Lorentz factor is possible, but the results of those simulations are difficult to scale back to the realistic parameters \citep{Sironi_2014,2017A&A...607A.112R}.

In this work, we focus on the nonlinear feedback of the instability on the beam and so we consider only the linear phase of the instability growth. We discuss in section \ref{sec:4} that under the instability feedback on the beam, the total electric field energy density stays always below the non-linear thresholds for the beam density we consider. In the next section, we look at the Fokker-Planck diffusion equation that describes the feedback of the electrostatic waves on the beam during the quasilinear regime. 

\subsection{Fokker-Planck diffusion equation for the pair beam}\label{sec:3.3}

The quasilinear regime is applicable when the total wave energy density is much smaller than that of the plasma. In this regime, the feedback of the electrostatic unstable waves on the beam is governed by the following Fokker-Planck diffusion equation \citep{Brejzman_1974}
\begin{equation}\label{eq:diff}
\begin{split}
    \frac{\partial f(p,\theta)}{\partial t} = & \frac{1}{p^2\theta}\frac{\partial}{\partial \theta}\left(\theta D_{\theta\theta} \frac{\partial f}{\partial \theta}\right) +  \frac{1}{p\theta}\frac{\partial}{\partial \theta}\left(\theta D_{\theta p} \frac{\partial f}{\partial  p}\right) \\ & + \frac{1}{p^2}\frac{\partial}{\partial p}\left(p D_{p \theta} \frac{\partial f}{\partial \theta}\right) + \frac{1}{p^2}\frac{\partial}{\partial p}\left(p^2 D_{p p} \frac{\partial f}{\partial p}\right), 
\end{split}
\end{equation}
where the diffusion coefficients are defined by the following resonance integrals \citep{1971JETP...32.1134R}
\begin{equation}
    D_{ij}(\boldsymbol{p}) = \pi e^2 \int d^3\boldsymbol{k} W(\boldsymbol{k}) \frac{k_i k_j}{k^2} \delta(\boldsymbol{k}\cdot\boldsymbol{v}-\omega_p),
\end{equation}
where the electric charge, $e$, is given in cgs units. The pair-beam distribution function, $f$, is given in spherical coordinates $(p,\theta,\varphi)$, and so is the wave-vector $\boldsymbol{k}$  $(k,\theta',\varphi')$. The angles $\theta$ and $\theta'$ are defined with respect to the beam propagation direction ($z-\rm{axis}$). Due to the azimuthal symmetry of the pair-beam distribution function, we can set $\varphi =0 $ and integrate over $\varphi'$, yielding (see appendix \ref{app:D})
\begin{equation}\label{eq:D}
\begin{split}
            \left\{\begin{array}{lr}
        D_{pp} \\
        D_{p\theta} \\
        D_{\theta \theta} 
        \end{array}\right\} = &  \frac{\pi m_e \omega_p^2}{n_e} \int_{\omega_p/c}^\infty k^2dk \int_{\cos{\theta_1'}}^{\cos{\theta_2'}} d\cos{\theta'} \frac{W(\boldsymbol{k}) }{kv_b} \\ &  \times \frac{1}{\sqrt{(\cos{\theta'}-\cos{\theta_1'})(\cos{\theta_2'}-\cos{\theta'})}} 
        \left\{\begin{array}{lr}
        1 \\
        \xi \\
        \xi^2 
        \end{array}\right\},
\end{split}
\end{equation}
where
\begin{equation}
    \xi = \frac{\cos{\theta}\frac{\omega_p}{kv_b}-\cos{\theta'}}{\sin{\theta}},
\end{equation}
and $v_b =c (1-\frac{1}{2\gamma^2})$ is the particle speed for Lorentz factor $\gamma$. The boundaries of the $\cos{\theta'}$ integration are fixed by the resonance condition
\begin{equation}
    \cos{\theta_{1,2}'} = \frac{\omega_p}{kv_b}\left[\cos{\theta}\pm\sin{\theta}\sqrt{\left(\frac{kv_b}{\omega_p}\right)^2-1}\right].
\end{equation}
The integrands are largest at the peak of the wave spectrum, therefore a proper numerical resolution of the spectrum is necessary when calculating the diffusion coefficients. We have changed the integration variables in appendix \ref{app:D} arriving at the coordinates $(k_\perp,\theta^{R})$ with $\theta^{R} = \left(\frac{ck_{||}}{\omega_p}-1\right)/(ck_\perp/\omega_p)$, for which the peak of the unstable modes is numerically well resolved and the diffusion coefficients are well defined by eq.\ref{eq:DFinal}. We also found that the Lorentz factor, $\gamma$ dependence of the diffusion coefficients is negligible compared to the beam angle, $\theta$.

In the next section, we describe the numerical setup of our study of the instability feedback and present our results. 

\section{Numerical results}\label{sec:4}

We have calculated the rate of change for every term on the right-hand side of the Fokker-Planck equation (eq.\ref{eq:diff}), using the diffusion coefficients of a wave spectrum generated by the growth rate presented in section \ref{sec:3.1}. We found that the diffusion term $D_{\theta \theta}$ exceeds the other terms by orders of magnitude in the phase-space region containing the bulk of the beam particles. Evaluating the maximum rate across the entire parameter space for every term, we found the following ratio between the different terms: $\theta \theta: \theta p: p \theta: pp \approx 1:10^{-3}:10^{-5}:10^{-8}$. 

Given this result, we initially neglect all the subdominant terms and in section \ref{sec:4.1} consider only the diffusion term $D_{\theta \theta}$. We will check the validity of this approximation in section \ref{sec:4.2} as we analyse the effect of the subdominant terms as the $\theta \theta$ diffusion modifies the beam. We also analyse the dependence of our results on the beam parameters in section \ref{sec:4.3}. Finally, we add the continuous pair production to our simulation setup in section \ref{sec:4.4}.

\begin{figure}
\centering
        \includegraphics[width=\columnwidth]{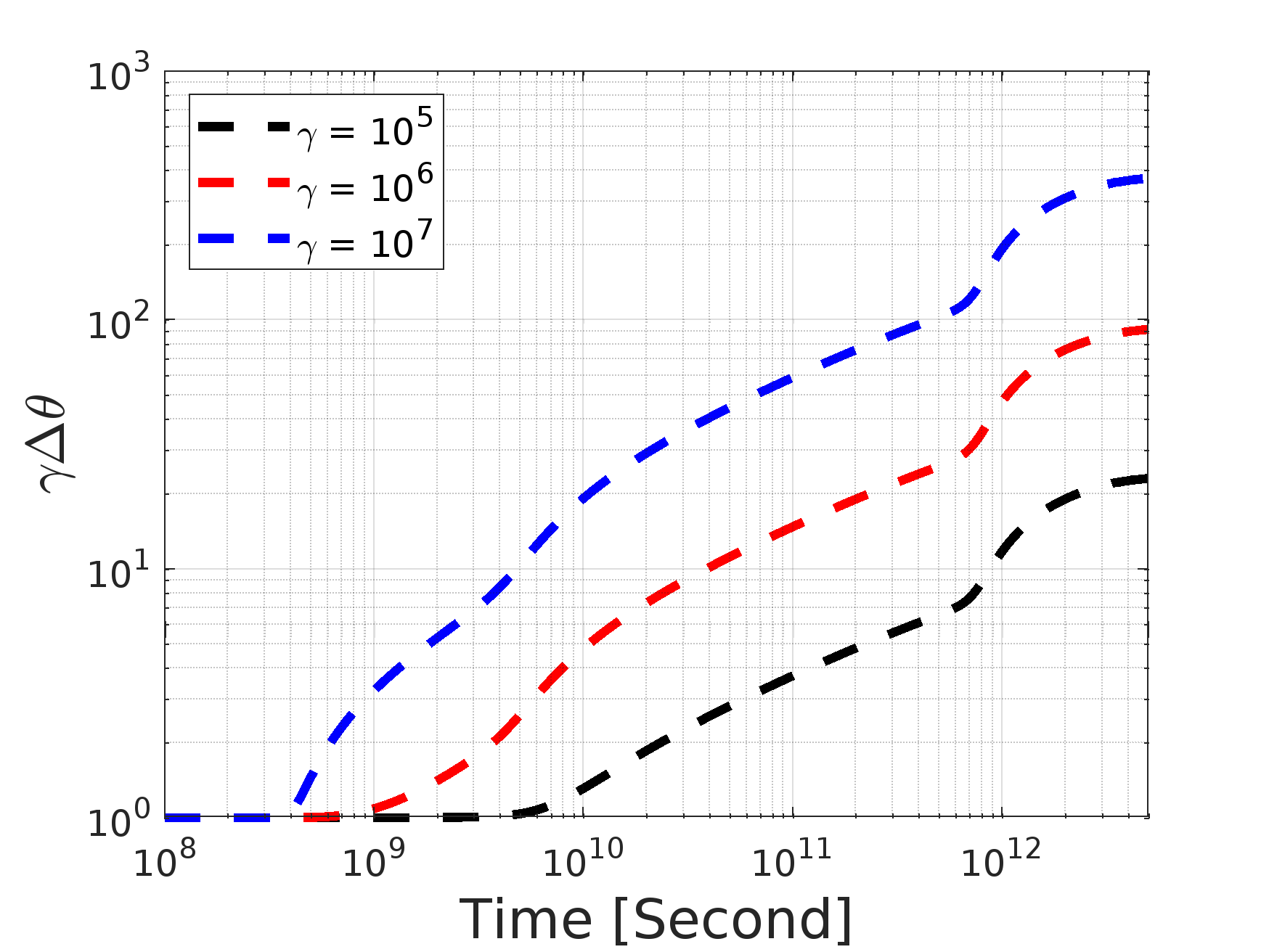}
        \caption{The angular spread for different Lorentz factors of the beam as a function of time during the angular diffusion feedback simulation presented in section \ref{sec:4.1}. The turn-up around the time of $7 \times 10^{11}$ seconds is due to the growth of the wave spectrum's third peak as seen in Fig.\ref{fig:lnWKpar}.}
        \label{fig:deltatheta}
\end{figure}

\begin{figure}
\centering
        \includegraphics[width=\columnwidth]{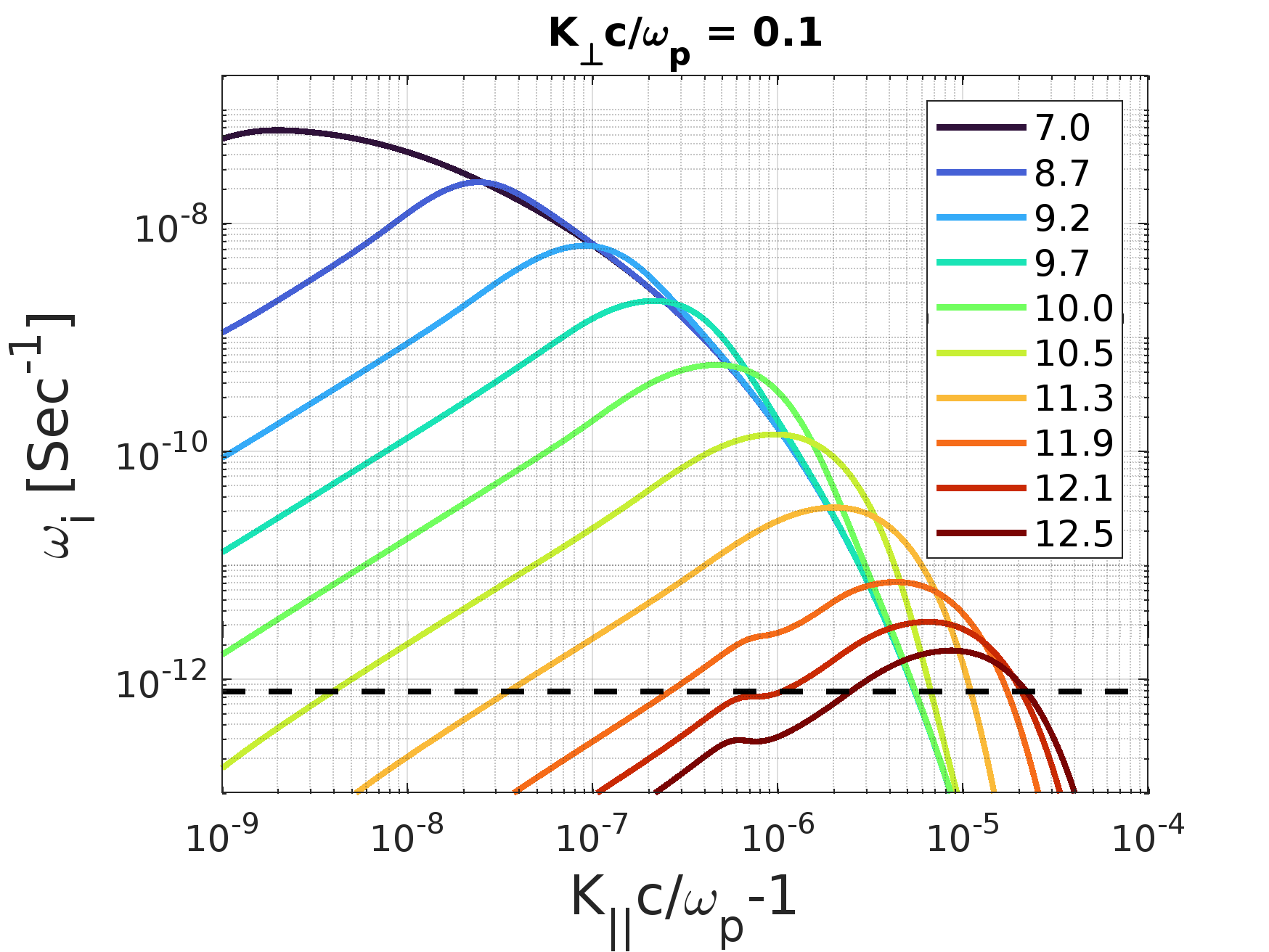}
        \caption{Evolution of the linear growth rate of the instability for a fixed perpendicular wave number during the angular diffusion feedback simulation presented in section \ref{sec:4.1}. The black dashed line represents the collisional damping rate. Legend values are common logarithms of time in seconds. Throughout the simulation, we observed that the linear growth rate has maintained its initial profile with perpendicular wave numbers as in Fig.\ref{fig:wi}. 
        }
        \label{fig:wiKpar}
\end{figure}

\subsection{Simulation of the $\theta \theta$ angular diffusion feedback} \label{sec:4.1}

Having established that $D_{\theta\theta}$ initially dominates over the other diffusion terms by orders of magnitudes, we perform here a numerical simulation of instability feedback including only this term. We introduce the simulation setup in section \ref{sec:4.1.1} and present the results in section \ref{sec:4.1.2}.

\subsubsection{Simulation setup} \label{sec:4.1.1}

The first numerical simulation of the beam-plasma system only includes the first term on the right-hand side of eq.\ref{eq:diff}
\begin{equation}\label{eq:1Ddiff}
     \frac{\partial f(p,\theta)}{\partial t} = \frac{1}{p^2\theta}\frac{\partial}{\partial \theta}\left(\theta D_{\theta\theta} \frac{\partial f(p,\theta)}{\partial \theta}\right),
\end{equation}
coupled time-dependently with the waves' spectral evolution equation (eq.\ref{eq:W}). The linear growth rate of the instability (eq.\ref{eq:growth}) and the diffusion coefficients (eq.\ref{eq:D}) involve integration over the beam distribution function and the wave spectrum, respectively.

We solve eq.\ref{eq:1Ddiff} using the Crank–Nicolson scheme along with the FTCS scheme for the wave equation, eq.\ref{eq:W}. We used a dynamical time step of $\omega_{i,\text{max}}^{-1}$ as the default time step with an upper limit set by the fastest rate of change of the distribution. We tested this by using time steps that are 10 times smaller. In order to properly resolve the narrow wave spectrum we use a logarithmic grid in the coordinates $(ck_\perp/\omega_p,\theta^R)$ where $\theta^R = (ck_{||}/\omega_p-1)/(ck_\perp/\omega_p)$. We verified convergence in our grid resolution for both the wave spectrum and the beam distribution. The initial beam distribution is as described in section \ref{sec:2}, and the initial wave energy density corresponds to the fluctuation level \citep{Vafin_2019}.

\begin{figure}
\centering
        \includegraphics[width=\columnwidth]{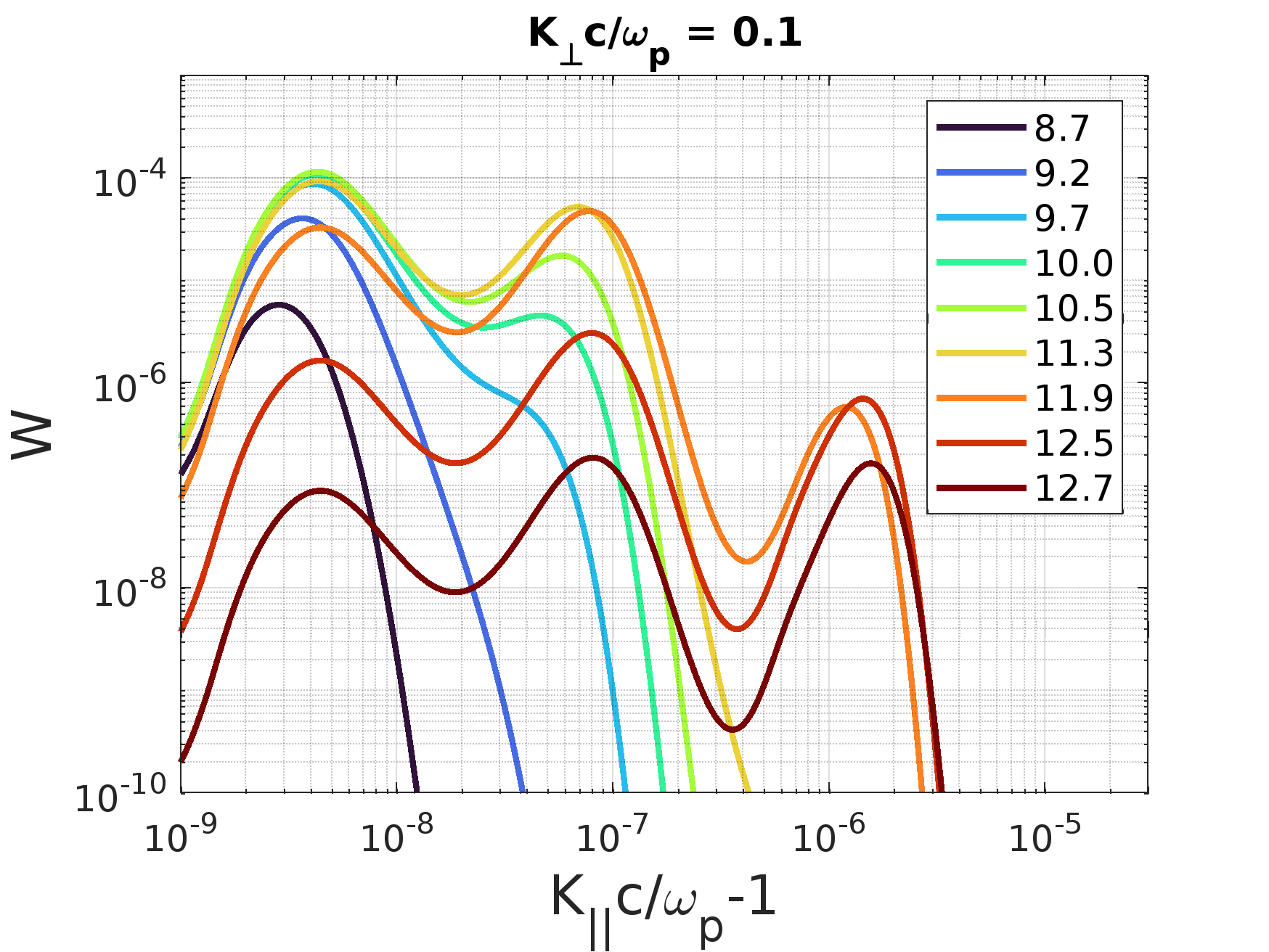}
        \caption{The time evolution of the wave spectrum for fixed perpendicular wave number for the angular diffusion feedback simulation presented in section \ref{sec:4.1}. The formation of the peaks here is due to the spectral change of the linear growth rate with time as shown in Fig.\ref{fig:wiKpar}. Legend values are common logarithms of time in seconds.}
        \label{fig:lnWKpar}
\end{figure}

\begin{figure}
\centering
        \includegraphics[width=\columnwidth]{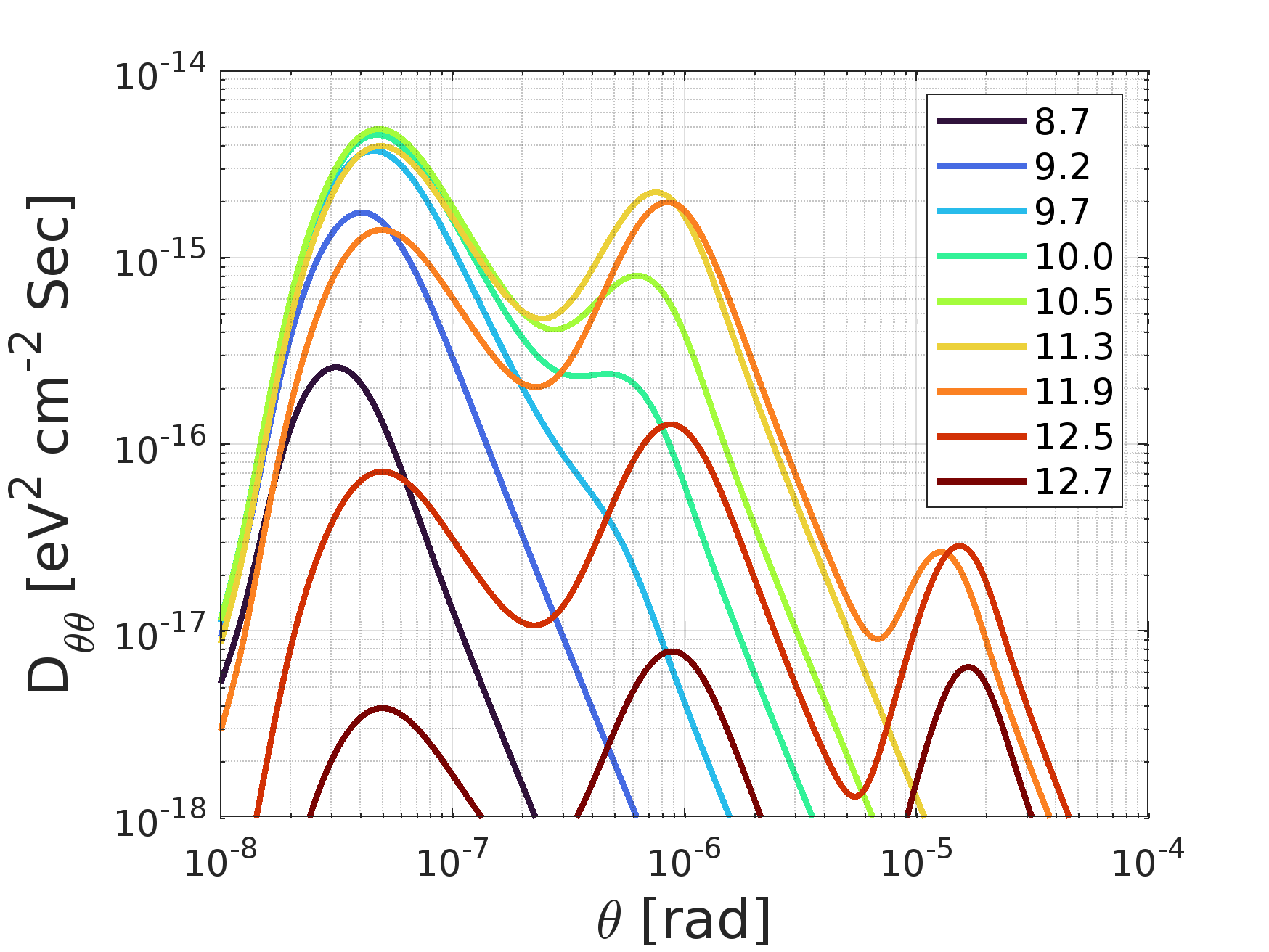}
        \caption{The $D_{\theta\theta}$ for $\gamma=10^6$ as a function of time in the angular diffusion feedback simulation presented in section \ref{sec:4.1}. Legend values are common logarithms of time in seconds.}
        \label{fig:Dt}
\end{figure}

\subsubsection{Results} \label{sec:4.1.2}

We found that the instability feedback severely increased the beam's angular spread. This broadening strongly depends on the Lorentz factor of the beam particles. In Fig.\ref{fig:deltatheta}, we show the angular spread for different beam Lorentz factors. We see that particles with larger Lorentz factors get scattered earlier since those particles are in resonance with faster-growing wave modes, and so the scattering feedback affects them earlier.

The angular spreading of the beam immediately shifts the resonant wave numbers. In Fig.\ref{fig:wiKpar}, we see the reduction of the growth rate for the parallel wave numbers during the simulation time for a fixed perpendicular wave number of 0.1. This reduction starts at the fastest growing modes as they quickly scatter their resonant particles, and with time it extends to slower growing modes at higher parallel wave numbers. We found that the initial profile of the linear growth with respect to the perpendicular wave numbers, as shown in Fig.\ref{fig:wi}, doesn't change during the time evolution. 

The resulting wave spectrum of the time-dependent linear growth rate is shown in Fig.\ref{fig:lnWKpar}. In the beginning, the fastest-growing modes form a spectral peak. Once the beam widens, the slower modes at higher parallel wave numbers start forming a second peak until the wave's intensity is sufficient to kick the beam particles to higher angles. The process keeps repeating until the linear growth rate becomes less than or comparable to the collisional damping rate (presented by the dashed black line in Fig.\ref{fig:wiKpar}). By the time we stop the simulation at $5\times 10^{12}$ seconds, all modes are collisionally damped.

In Fig.\ref{fig:Dt}, we show the diffusion coefficient, $D_{\theta\theta}$, at various times. The variation of the diffusion coefficient with beam angle, $\theta$, closely resembles that of the wave spectrum with parallel wave numbers, as represented in Fig.\ref{fig:lnWKpar}. This is due to the resonance relation between the beam angle and the parallel wave numbers ($\theta = (ck_{||}/\omega_p-1)/(ck_\perp/\omega_p)$ in the regime $\left(ck_\perp/\omega_p\right)^2 >> \left(\frac{ck_{||}}{\omega_p}-1\right)$). 

Peaks in the wave spectrum (Fig.\ref{fig:lnWKpar}) and diffusion coefficient (Fig.\ref{fig:Dt}) arise from the evolving linear growth rate (Fig.\ref{fig:wiKpar}). The initial peak forms at wave numbers with the highest growth rates, causing resonant particles to diffuse, smoothly shifting the peak. A second peak emerges as particles diffuse further, resonating with higher wave numbers. Eventually, collisional damping leads to the decay of both the first and second peaks. A third peak follows in the same mechanism, decaying at a lower amplitude due to the fall of the growth rate below the collisional damping shortly.

The appearance of the last peak in the diffusion coefficient profile, manifesting at angular values around $10^{-5}$ radians, causes the observed surge in the beam angular spread after $7 \times 10^{11}$ seconds (Fig.\ref{fig:deltatheta}). The time scale of this upturn in angular spread is governed by the interplay between the evolving linear growth rate and the collisional damping rate. We also see that around this time pairs with different Lorentz factors react to the same resonant unstable modes, because in the regime $\left(ck_\perp/\omega_p\right)^2 >> \left(\frac{ck_{||}}{\omega_p}-1\right)$ waves with a certain parallel wave number are resonant with the beam particles at a certain angle, whatever their momentum.

We see in Fig.\ref{fig:deltatheta} that by the time the instability has saturated, the angular spread of pairs with Lorentz factor $10^6$ has increased by around two orders of magnitudes, much more than the factor of ten reported by \citet{Perry_2021}. The main reasons for this higher spread are the smaller collisional damping rate and the higher beam density we used. 

We found that during the entire simulation time, the wave energy density never exceeded $10^{-3}$ of the beam energy density. This level of the wave intensity is lower than that needed for efficient operation of nonlinear Landau damping and the Modulation instability \citep{Vafin_2019,Chang_2014,Miniati_2013}. Therefore, the effect of these non-linear processes on the instability development might be minimal compared to that of the diffusive feedback on the beam.

\begin{figure}
\centering
        \includegraphics[width=\columnwidth]{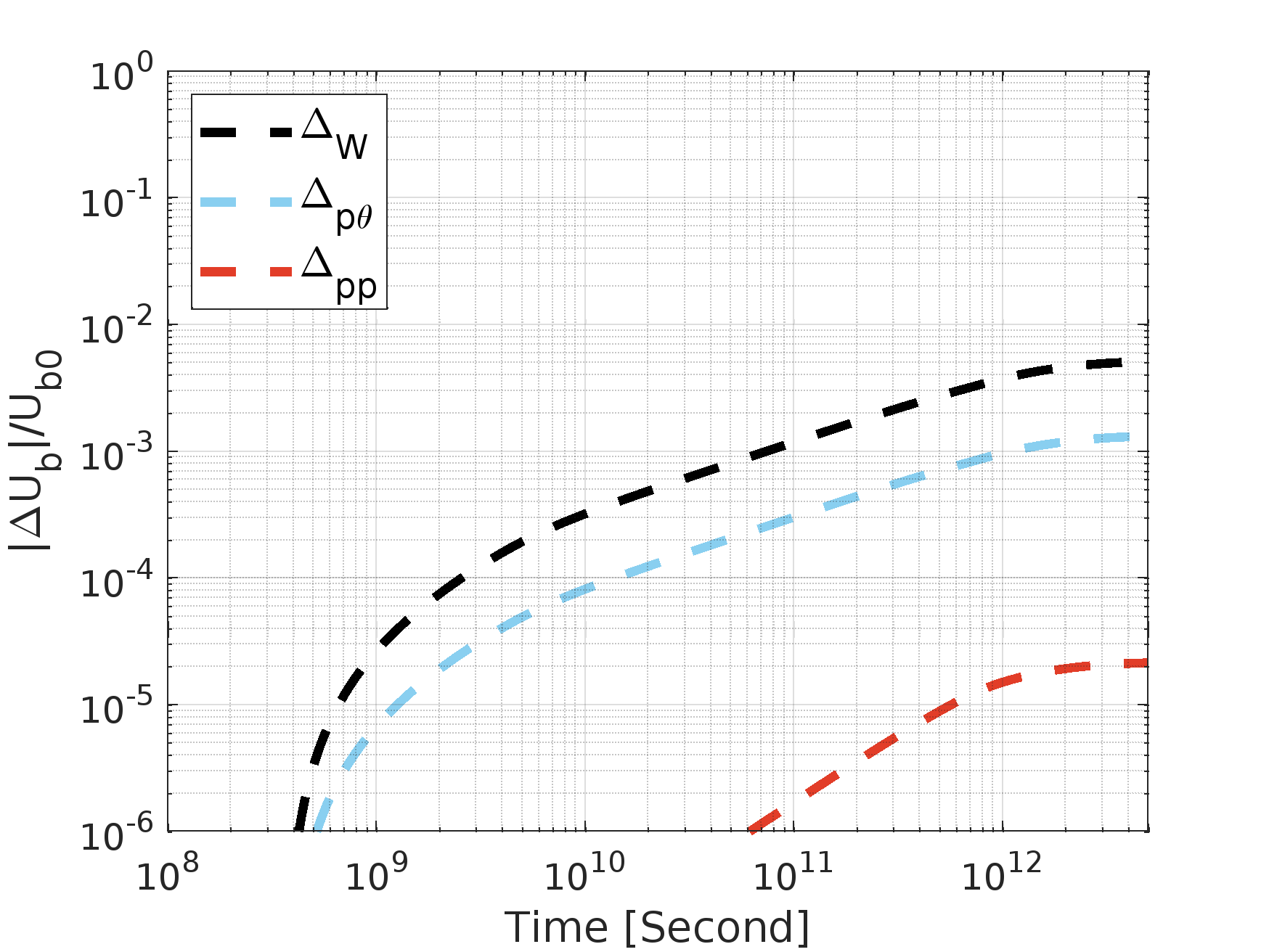}
        \caption{The accumulated change in the beam energy during the angular diffusion feedback simulation presented in section \ref{sec:4.1}. The black dashed line ($\Delta_W$) represents the beam energy fraction going into unstable wave growth. The dashed cyan line ($\Delta_{p\theta}$) and the dashed red line ($\Delta_{pp}$) represent the fraction of the beam energy loss and gain due to the momentum diffusion by the $p\theta$ and the $pp$ terms, respectively.}
        \label{fig:timeintegral}
\end{figure}

We also calculated the total energy transferred from the beam to the waves by integrating the energy loss rate of the beam given in eq.\ref{eq:lossrate} over time. The result is given by the black dashed line in Fig.\ref{fig:timeintegral}. We see that the beam lost less than $1\%$ of its total initial energy by the time the instability development was saturated by the widening feedback. Those results suggest that the feedback widening severely limits the energy transfer from the beam to the waves. We explore whether this situation changes as we use different beam densities in section \ref{sec:4.3}.

Up to here, we only included the initially dominant term $D_{\theta \theta}$ of the right-hand side of eq.\ref{eq:diff}. In the next section, we analyse the feedback of the other subdominant terms as the dominant $\theta \theta$ diffusion widens the pair beam.

\begin{figure}
\centering
        \includegraphics[width=\columnwidth]{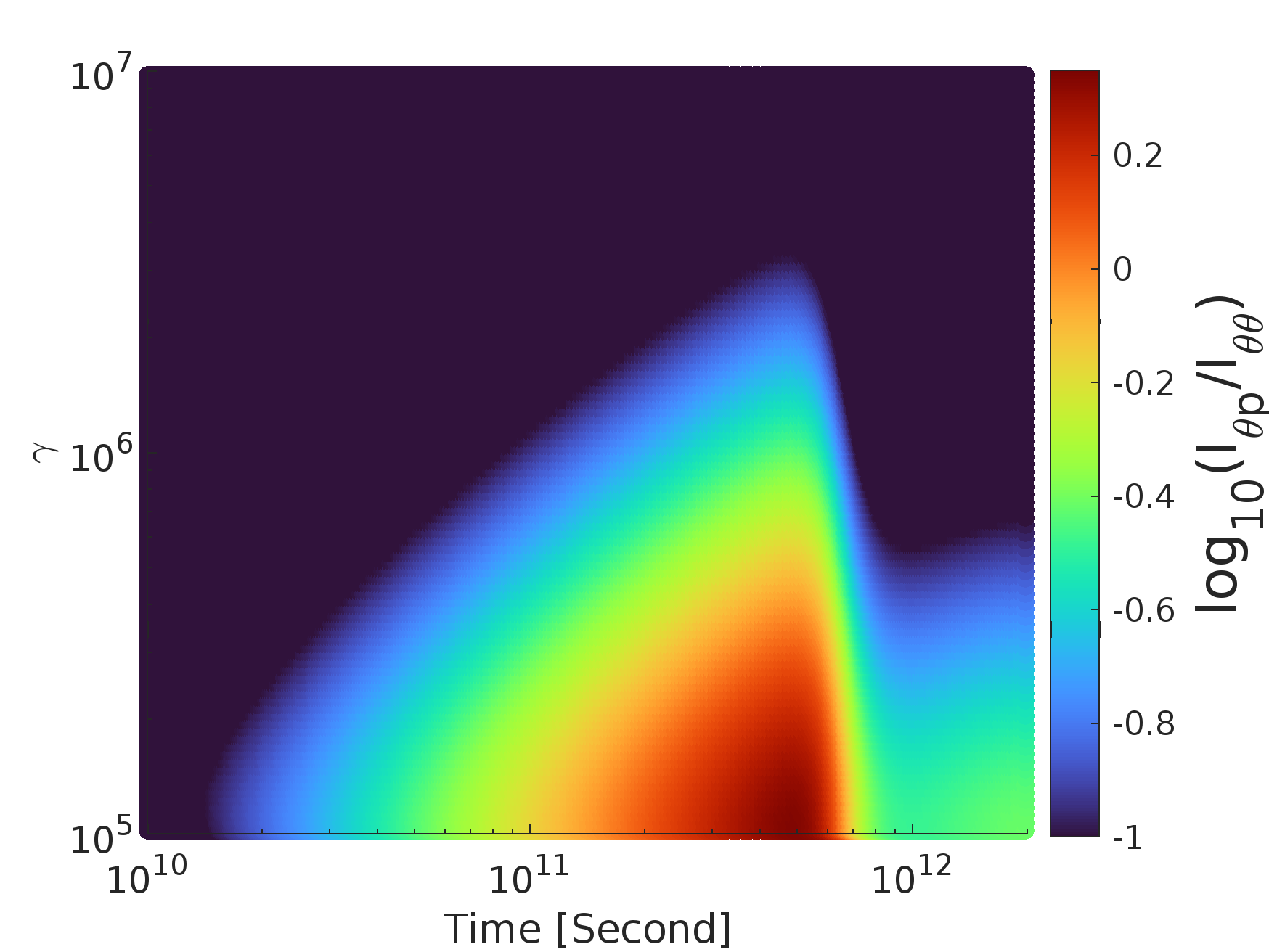}
        \caption{The logarithm of the ratio of $I_{\theta p}$ (eq.\ref{eq:tp}) and $I_{\theta \theta}$ (eq.\ref{eq:tt}). The diffusion $\theta p$ dominates over the term ($\theta \theta$) in the orange and red areas with values higher than zero, while it contributes less than 10$\%$ in the dark blue area. The drop in the ratio just before the rim at $10^{12}$ seconds is due to the increase in the widening as a result of wave growth outside the initial resonance region. After $10^{12}$ seconds, the collisional damping effectively damps the waves, and the impact of both terms declines.}
        \label{fig:ItpItt}
\end{figure}

\subsection{2D analysis of the diffusion equation} \label{sec:4.2}

We analyse here the effect of the subdominant terms as the beam widens, using the time-dependent beam distribution that we numerically derived and discussed in the previous section.

For the momentum diffusion of the beam (third and fourth terms on the RHS of eq.\ref{eq:diff}), we can calculate the energy loss or gain rate of the beam by inserting the corresponding time derivative of the beam distribution eq.\ref{eq:diff} in the total rate of change of the beam energy. After integrating by parts we get the following relation for $p\theta$ diffusion

\begin{equation} \label{eq:pt}
\begin{split}
    \frac{dU_b}{dt}\Bigr|_{p \theta} (t) & = 2 \pi m_e c^2  \int d\theta \theta \int dp p^2  \gamma  \frac{df}{dt}\Bigr|_{p\theta}(p,\theta,t) \\ 
    &  = - 2 \pi c  \int d\theta \theta \int dp  p D_{p\theta} \frac{\partial f}{\partial \theta} (p,\theta,t),
    \end{split}
\end{equation}
and the following for $pp$ diffusion
\begin{equation}\label{eq:pp}
    \frac{dU_b}{dt}\Bigr|_{pp} (t) = - 2 \pi c  \int d\theta \theta \int dp  p^2 D_{pp} \frac{\partial f}{\partial p}(p,\theta,t).
\end{equation}

Looking at the overall sign of eq.\ref{eq:pt}, we see that the diffusion $p\theta$ involves a global energy loss of the beam since $D_{p\theta}$ and the angular derivative are always negative. We also found that the dominant feedback of the diffusion $pp$ is an energy gain of the beam, as $D_{pp}$ is always positive and the beam distribution function declines for $\gamma \gtrsim 10^5$ (see Fig.\ref{fig:fbg}).

Integrating eq.\ref{eq:pt} and eq.\ref{eq:pp} over the simulation time and dividing by the total beam initial energy, we see in Fig.\ref{fig:timeintegral} the accumulated fraction of the beam energy lost and gained. We observe that $p\theta$ diffusion could eliminate only around 0.1$\%$ of the beam total energy by the end of the simulation whereas $pp$ diffusion increases it by a negligible fraction. Therefore, it is evident that the cumulative effect of the diffusive momentum flux on the beam is insignificant compared to the scattering. 

Now, we proceed to the analysis of the second term on the RHS of eq.\ref{eq:diff}, $\theta p$ diffusion. This diffusion involves angular flux as the $\theta \theta$ diffusion, but it can result in both the narrowing and widening of the beam depending on the beam momentum gradient. Pairs with negative momentum gradient, $\gamma>10^5$, experience narrowing whereas the ones with positive momentum gradient, $\gamma<10^5$, experience a widening. In Fig.\ref{fig:ItpItt}, we have compared the normalized angular integral of the absolute rate of change of the $\theta \theta$ diffusion for a certain beam Lorentz factor 
\begin{equation}\label{eq:tt}
     I_{\theta \theta} = \int d\cos{\theta} \left|\frac{df}{dt}\Bigr|_{\theta\theta}\right|  = \int d\cos{\theta} \left|\frac{1}{p^2\theta}\frac{\partial}{\partial \theta}\left(\theta D_{\theta\theta} \frac{\partial f}{\partial \theta}\right)\right|,
\end{equation}
with that one of the $\theta p$ diffusion
\begin{equation}\label{eq:tp}
     I_{\theta p} =\int d\cos{\theta} \left|\frac{df}{dt}\Bigr|_{\theta p}\right| = \int d\cos{\theta} \left|\frac{1}{p\theta}\frac{\partial}{\partial \theta}\left(\theta D_{\theta p} \frac{\partial f}{\partial  p}\right)\right|.
\end{equation}

It is noticeable in Fig.\ref{fig:ItpItt} that the ratio of $I_{\theta p}$ and $I_{\theta \theta}$ increases gradually until it drops after $7\times 10^{11}\ \mathrm{s}$. The reason for the increase is that the diffusive flux of $\theta \theta$ decreases as the beam profile flattens, while the diffusive flux of $\theta p$ remains relatively constant as the momentum gradients are not impacted by the beam broadening. The drop after $7\times 10^{11}$ seconds is due to the increase of diffusion $\theta \theta$ by the accumulated wave density outside the initial resonance region that we discussed in the previous section. 

In Fig.\ref{fig:ItpItt} we see that the $\theta p$ diffusion becomes dominant for Lorentz factors less than $10^6$ at times much earlier than their inverse Compton cooling time ($\gtrsim 10^{13}\ \mathrm{s}$). For these particles including this diffusion is necessary. However, there is a minimal impact of the $\theta p$ diffusion on the pairs that are capable of giving IC emission in the detectable GeV band (Lorentz factors of $10^6$ or slightly higher). This indicates that $\theta p$ diffusion might not impact the GeV-scale cascade emission as strongly as the $\theta \theta$ diffusion does.

\subsection{Parameters dependence}\label{sec:4.3}

\begin{figure}
\centering
        \includegraphics[width=\columnwidth]{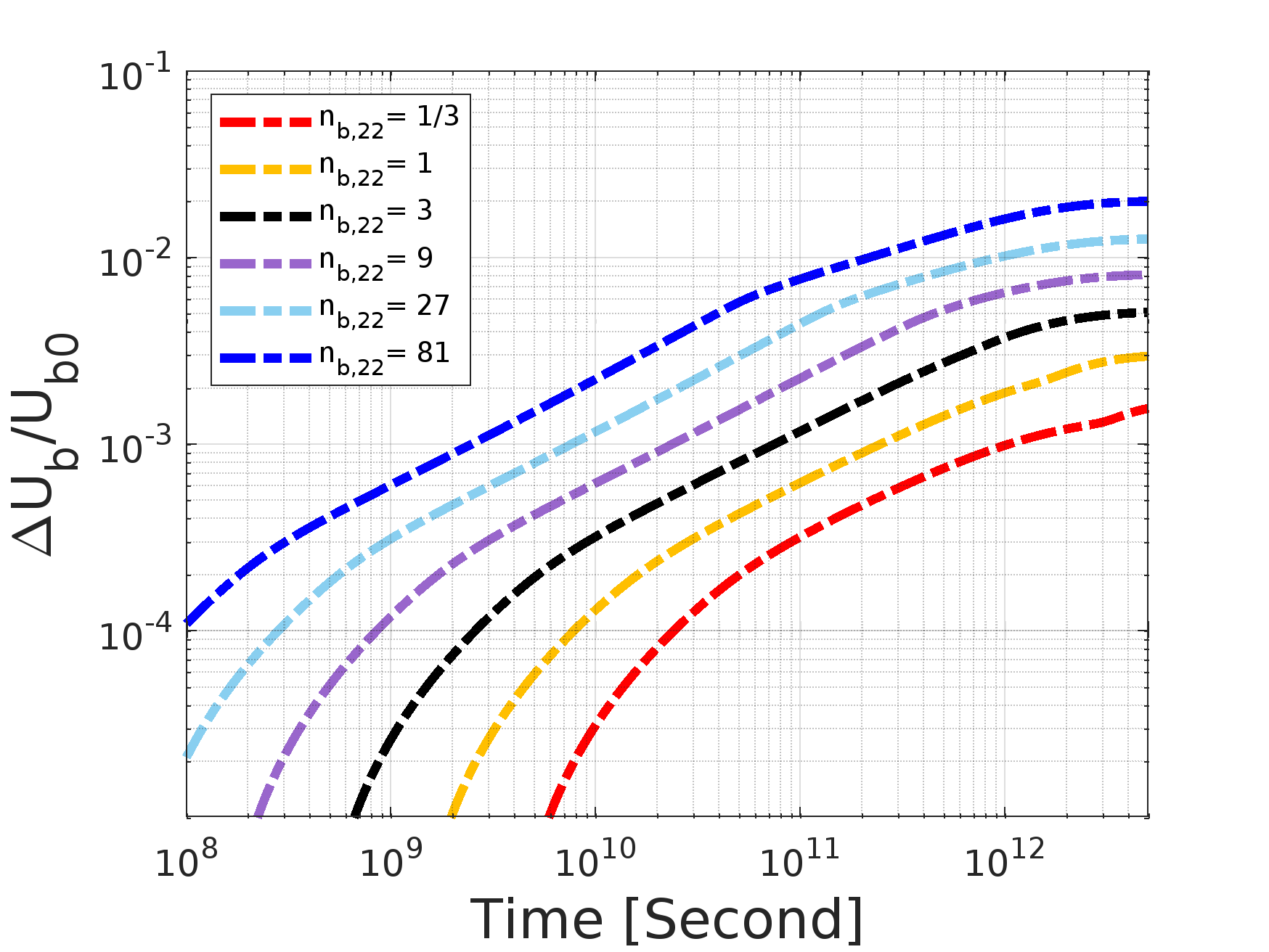}
        \caption{The accumulated fraction of the beam energy lost as a function of time due to the wave growth during the angular diffusion simulations feedback with different values of the beam density. All the values are in units of $10^{-22}$ cm$^{-3}$.}
        \label{fig:nb_loss}
\end{figure}

In the simulation discussed in section \ref{sec:4.1}, we used a fiducial pair beam density at a distance of 50 Mpc from the blazar, $3\times 10^{-22}$cm$^{-3}$ \citep{Vafin_2018}. However, the beam density changes under different conditions, such as varying the distance from the source, changing the source's luminosity, or using different EBL models in the calculations. Here we vary the beam density using the same setup as in section \ref{sec:4.1} and investigate its impact on our results.

In Fig.\ref{fig:nb_loss}, we see the fraction of the beam energy lost by the instability for different beam densities. As the beam density is increased, the instability develops earlier and takes more energy from the beam. However, the beam lost only 2$\%$ even for a very high beam density, $8 \times 10^{-21}\ $cm$^{-3}$. Therefore, the fundamental physical behaviour of the system remains consistent, the beam experiences expansion with a negligible energy loss of its initial energy as the instability is saturated by the beam expansion. 

We noticed that the wave intensity during the simulations with beam density higher than $10^{-21}$cm$^{-3}$ has exceeded the threshold for the non-linear modulation instability, which we didn't include in our calculations. Those processes will impose further restrictions on the growth of the unstable modes.

We observed that the angular spread increased by a factor of $1.5$ when the beam density was inflated by a factor of three. This scaling can be attributed to the fact that the linear growth rate is linearly proportional to the pair beam density and inversely proportional to the square of the beam angular spread, $\omega_i \propto \frac{n_b}{\Delta\theta^2}$. Therefore, increasing the beam density by a factor $C$ requires an increase in the angular spread by a factor of approximately $\sqrt{C}$ to maintain the reduction of the linear growth rate to the collisional damping rate at the time when the instability has saturated.

In the remainder of this section, we will discuss the influence of the cut-off energy in the intrinsic gamma-ray spectrum on the results of section \ref{sec:4.1}. \citet{Vafin_2018} used an intrinsic power-law gamma-ray spectrum with a step function cut-off at the energy of 50 TeV. However, in the end, they used the attenuated gamma-ray spectrum at a distance of 50 Mpc to calculate the accumulated pair beam spectrum over a certain path length. At a distance of 50 Mpc from the blazar, the majority of gamma rays with energies higher than 10 TeV have already been absorbed. The mean free path for a gamma-ray with energy $E_\gamma$ for pair production with the EBL photons is given by 
\begin{equation}
    \lambda_{\gamma \gamma} \approx 80 \left(1+z\right)^{-\xi} \left(\frac{E_\gamma}{10 \text{TeV}}\right)^{-1} \text{Mpc},
\end{equation}
where $\xi =4.5$ and $\xi =0$ for redshifts of $z\leq1$ and $z>1$, respectively \citep{refId0,PhysRevD.80.123012}. Therefore, any cut-off energy above the $10$-TeV threshold will have only a minimal impact.

%\begin{figure}
%\centering
%        \includegraphics[width=\columnwidth]{Fig/Inj_wiKpar.png}
%        \caption{The linear growth rate for parallel wave numbers during the simulation with continuous pair injection in section \ref{sec:4.4}. We see the linear growth rate eventually balance the collisional damping rate across the wave spectrum. We see also that the resonance region keeps expanding due to the ongoing expansion of the beam particles with higher opening angles.}
%        \label{fig:wiInj}
%\end{figure}

\subsection{Simulation with injection}\label{sec:4.4}

In section \ref{sec:4.1}, we found that the instability growth is severely reduced by beam broadening to the point that it cannot be isolated from the production and the cooling rates of the beam. In this section, we include the pair creation rate in the evolution equation of the beam.

The beam distribution found in \citet{Vafin_2018} was calculated as the accumulation of pairs over the path length of $7.7 \times 10^{12}$ light-seconds, using a constant production rate $Q_{ee}$. We added this production rate to the beam evolution equation along with the dominant $\theta\theta$ diffusion term,
\begin{equation}\label{eq:diffinj}
\begin{split}
    \frac{\partial f(p,\theta)}{\partial t} =  \frac{1}{p^2\theta}\frac{\partial}{\partial \theta}\left(\theta D_{\theta\theta} \frac{\partial f}{\partial \theta}\right) + Q_{ee}.
\end{split}
\end{equation}

Using the same simulation setup as described in section \ref{sec:4.1}, we numerically solved the coupled system of the evolution equations (eq.\ref{eq:diffinj} and eq.\ref{eq:W}). For times much less than $10^{13}$ seconds we found essentially the same behaviour of the system as without injection. After $10^{13}\ \mathrm{s}$, a new quasi-steady state of the beam distribution and the waves spectrum emerges. 

\begin{figure}
\centering
        \includegraphics[width=\columnwidth]{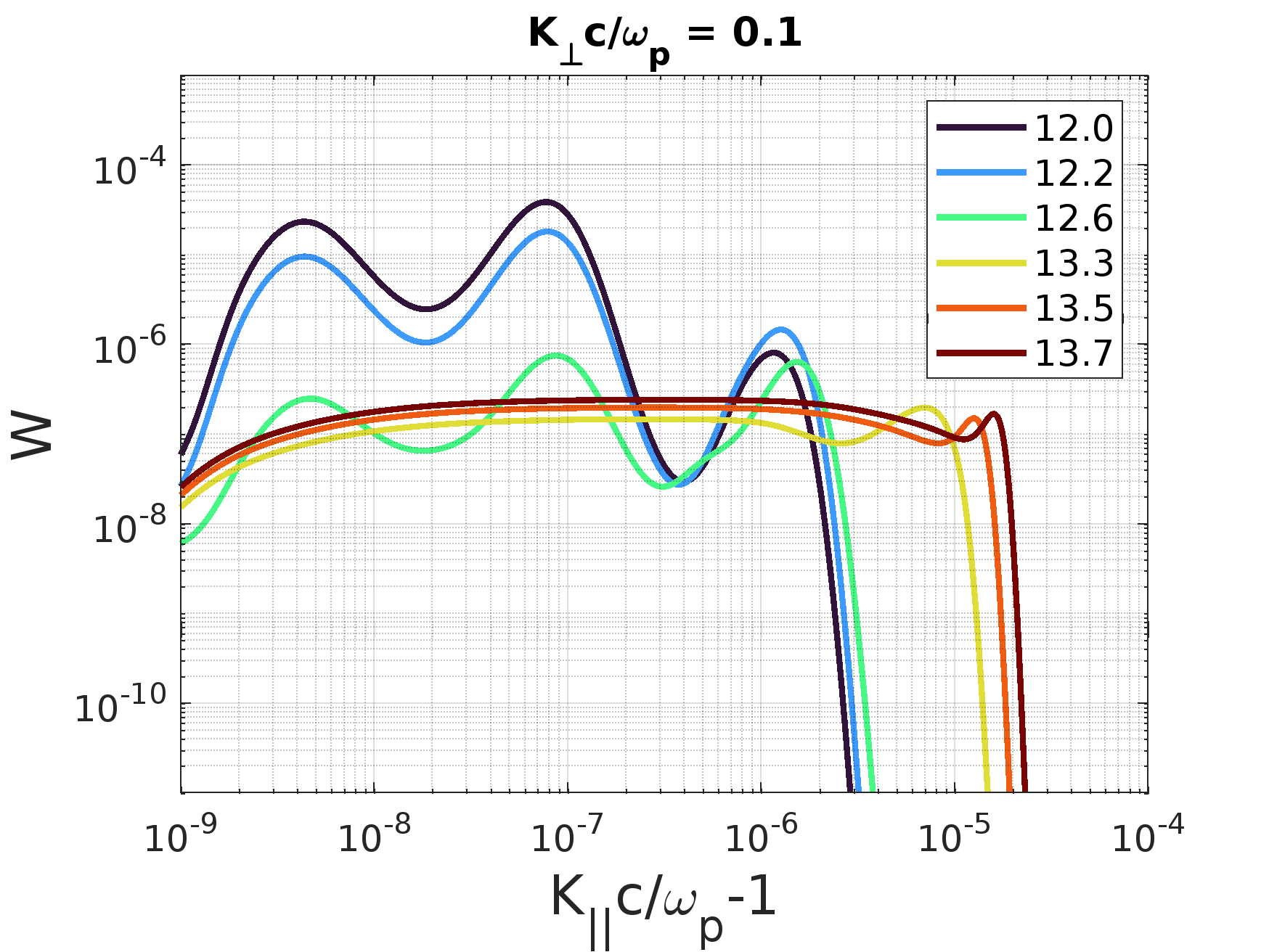}
        \caption{The wave spectrum as a function of time during the injection simulation presented in section \ref{sec:4.4}. Legend values are common logarithms of time in seconds. We see the steady-state wave spectrum emerging after $10^{13}\ \mathrm{s}$.}
        \label{fig:lnWInj}
\end{figure}

\begin{figure}
\centering
        \includegraphics[width=\columnwidth]{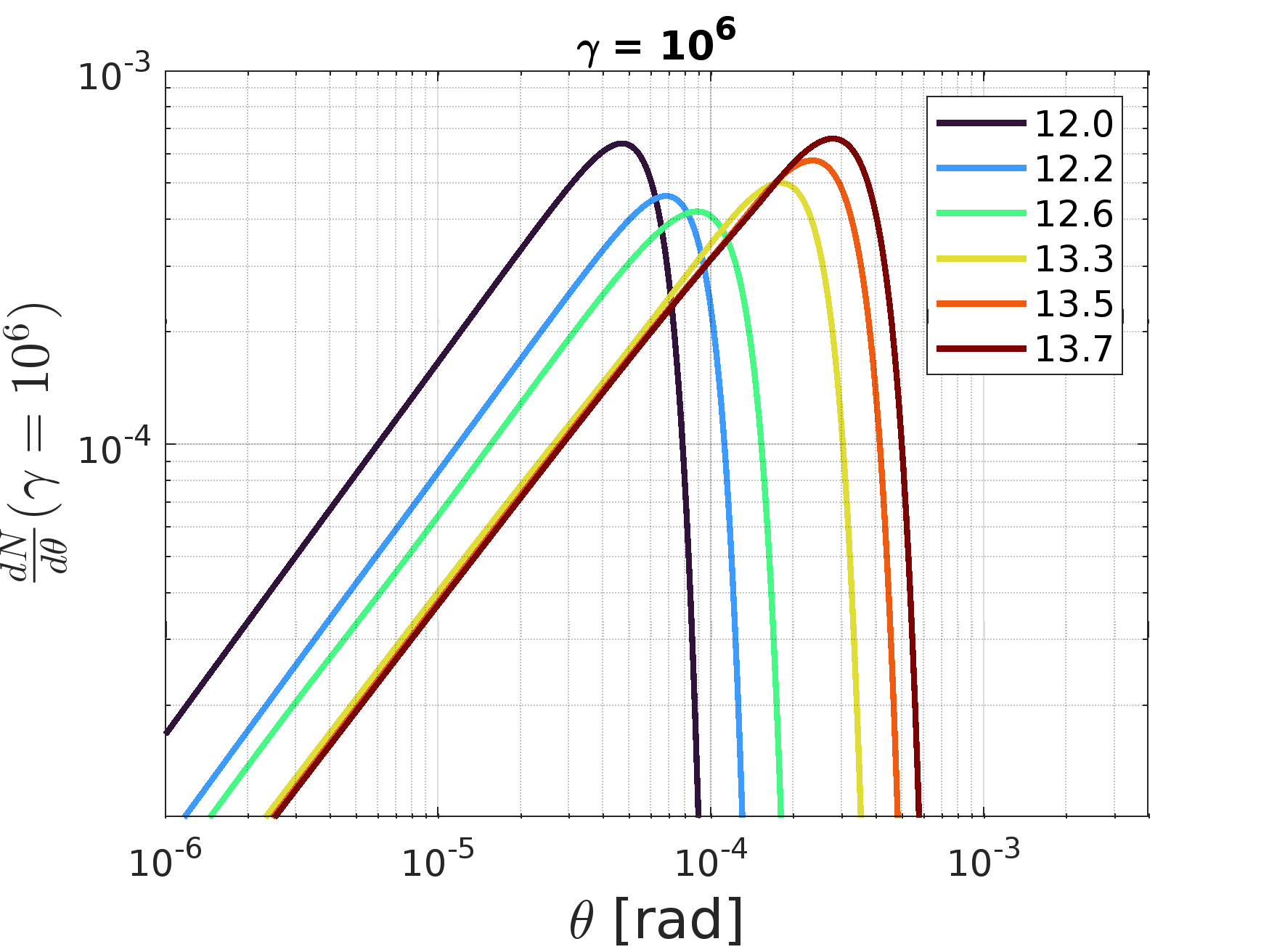}
        \caption{The angular profile of pairs with Lorentz factor of $10^6$ during the injection simulation presented in section \ref{sec:4.4}. Legend values are common logarithms of time in seconds. We see that around the IC cooling time of $5\times10^{13} \mathrm{s}$ for those pairs, they have been deflected by around $4\times 10^{-4}$ radians which yields a time delay of around 10 years for the GeV cascade.}
        \label{fig:Ng6Inj}
\end{figure}

The creation of highly focused pairs with beam angles of the order $\gamma^{-1}$ increases the linear growth rate at wave numbers in resonance with these particles. Ultimately, this leads to a quasi-equilibrium of the wave spectrum and the beam distribution. On the wave side, the linear growth rate and the collisional damping rate balance across the resonant wave numbers, resulting in a steady-state wave spectrum as shown in Fig.\ref{fig:lnWInj}. On the beam side, the diffusive scattering compensates the pair production, keeping the beam expanding as shown in the angular profile of pairs with a Lorentz factor of $10^6$ in Fig.\ref{fig:Ng6Inj}. This ongoing expansion of the beam extends the unstable modes to higher parallel wave numbers as shown in Fig.\ref{fig:lnWInj}

We have stopped the simulation after $5\times 10^{13}\ \mathrm{s}$, which corresponds to the IC cooling time of pairs with a Lorentz factor of $10^6$. By this time the pairs have experienced a diffusive deflection up to angles of around $4\times 10^{-4}$ radians. This deflection results in an arrival time delay of the secondary GeV-band photons emitted by those pairs \citep{2009PhRvD..80l3012N}. The arrival time delay of secondary gamma rays emitted by pairs that have undergone a deflection by an angle of $\Delta \theta$ from the primary gamma-ray propagation direction is given by the following formula 

\begin{equation}
    \Delta t_{\text{delay}} \simeq \frac{\Delta \theta^2}{2} \frac{D_c D_b - D_c^2}{c D_b},
\end{equation}
where $D_c$ is the distance between the emitting pairs and the blazar and $D_b$ is the distance between the blazar and the Earth. Given that for our simulation setup $D_c=50$ Mpc and $D_b=720$ Mpc for the fiducial $z=0.15$, the formula reduces to $\Delta t_{\text{delay}} = \Delta \theta^2 \times 7.6 \times 10^{7}$ years. Hence, the deflection of pairs with Lorentz factor of $10^6$ by $4\times 10^{-4}$ radians implies a time delay of around 10 years for the GeV-scale cascade emission produced at the distance 50 Mpc from the source. Calculating the deflection at different distances from the source is needed to find the impact on the observed cascade emission. This is beyond the scope of this paper and will be covered in future works.

As of the previous simulation in section \ref{sec:4.1}, we also found here that the wave energy density never exceeded  $10^{-3}$ of the total beam energy density, keeping the wave intensity at levels lower than what is needed for the non-linear processes to operate efficiently, again justifying their neglect.

This paper's calculations did not consider the IC scattering of the beam particles. For particles with Lorentz factors $\gamma < \frac{m_e c^2}{4 \epsilon_\text{CMB}} \sim 2.5 \times 10^8$, IC scattering occurs in the Thomson regime, leading to momentum loss without significant angular changes. While we simulate the beam until the IC cooling time of pairs with $\gamma = 10^6$, we anticipate a substantial decrease in the beam steady state for Lorentz factors above $10^6$. The steady-state wave spectrum is expected to undergo less significant changes however as it is influenced by the resonance condition that is set by the particles' angle irrespective of their energy. Nevertheless, the inclusion of IC cooling is crucial for a comprehensive understanding of the physical implications on the arrival time distribution of GeV-scale cascade emissions, and it is part of our future research plans.

\section{Conclusions}\label{sec:con}
We have explored the feedback of the electrostatic beam-plasma instability on blazar-induced pair beams. The feedback of the beam-plasma instability is described by a Fokker-Planck diffusion equation with diffusion coefficients that are dependent on the resonance condition between the unstable waves and the beam particles. This feedback is crucial for understanding the propagation of the blazar-induced pair beam in the IGM, in particular the question whether or not the instability is capable of draining the beam energy faster than inverse Compton cooling. Such insights hold significance in unravelling the underlying reasons for the absence of secondary GeV-scale emissions in several distant blazar spectra \citep{2009PhRvD..80l3012N, Broderick_2012}.

We solved the Fokker-Planck diffusion equation for the beam distribution function, coupled with the linear evolution equation of the plasma-wave spectrum. As the initial condition for the beam, we used the realistic two-dimensional beam distribution computed by \citet{Vafin_2018} for a distance of 50 Mpc from the blazar. Initially, the dominant feedback is angular broadening of the beam, stemming from the scattering of the beam particles by the excited waves. As the instability widens the beam, the instability growth rate is severely reduced, leading at the end to a negligible energy transfer from the beam to the plasma waves. These findings align with a recent study on instability feedback \citep{Perry_2021}.

Using the 2D time-dependent beam profile evolving by the predominant angular diffusion, we found that momentum diffusion does not have any significant impact on the beam. However, we found that another angular diffusion term, which is initially negligible, might become relevant and may narrow the beam particles with Lorentz factors between $10^5$ and $10^6$. Therefore, including this term in the feedback calculations is necessary for a comprehensive understanding of the instability impact on those pairs. However, the GeV-scale cascade is emitted by pairs with Lorentz factors of $10^6$ or slightly higher, and so the impact of this term on the GeV-scale secondary cascade might be limited.

In our analysis, we neglected non-linear wave interactions in the evolution of the wave spectrum. For beam density lower than $10^{-21}$ cm$^{-3}$, we found that the cumulative energy density of the electric field fluctuations remains below the critical thresholds required to trigger the significant impacts of the non-linear processes, such as non-linear Landau damping or the modulation instability. However, for higher beam densities the wave energy density exceeded the threshold for the non-linear modulation instability. Those non-linear processes would impose further restrictions on the growth of the unstable modes.

Lastly, we have included the continuous TeV pairs production in the Fokker-Planck diffusion equation. Unlike the previous simulation discussed in section \ref{sec:4.1}, in this particular configuration, the unstable modes do not decay after the beam has expanded but saturate at a finite amplitude. The wave spectrum reaches a quasi-equilibrium across the wave numbers resonant with the beam injection angles. The beam particles experience persistent scattering under the diffusive feedback of this steady-state wave spectrum. Then, beam particles with Lorentz factors of $10^6$ scatter up to angles of around $4\times10^{-4}$ radians within their IC cooling time. This results in a time delay of around 10 years in the arrival of the secondary GeV-scale cascade, assuming pairs at a distance of 50 Mpc from a blazar that is 720 Mpc away from Earth. We expect that this estimate depends on the beam density that varies along the propagation distance and with source luminosity. 

In the end, calculating the broadening at more points along the beam propagation is needed to understand the impact of the instability broadening on the GeV-scale cascade emission. Also, it’s essential to include the inverse Compton cooling in the beam distribution evolution equation to understand the long-term time evolution of the beam-wave system.

\section*{Acknowledgement}

This work was supported by the International Helmholtz-Weizmann Research School for Multimessenger Astronomy, largely funded through the Initiative and Networking Fund of the Helmholtz Association. We thank Andrew Taylor and Kfir Blum for the interesting discussion and helpful comments. MA thanks Karol Fulat for the interesting discussions.

%%%%%%%%%%%%%%%%% APPENDICES %%%%%%%%%%%%%%%%%%%%%

\appendix

\section{DIFFUSION COEFFICIENTS} \label{app:D}

The diffusion coefficients are given by 
\begin{equation}\label{eq:D1}
    D_{ij} = \pi e^2 \int d^3\boldsymbol{k} W(\boldsymbol{k},t) \frac{k_i k_j}{k^2} \delta(\boldsymbol{k}\cdot\boldsymbol{v}-\omega_p), 
\end{equation}
where the unstable wave wavevector $\boldsymbol{k}=(k,\theta',\varphi')$ and the beam particles velocity $\boldsymbol{v}=(v=c(1-\frac{1}{2\gamma^2}),\theta,\varphi=0)$ are both defined in the spherical coordinates with the beam propagation axis being the $z$-axis. Because of the azimuth symmetry, we set $\varphi=0$ without losing the generality yielding
\begin{equation}
    D_{ij}=\pi e^2 \int k^2 dk \int d\cos{\theta'} \int d\varphi' W(\boldsymbol{k},t) \frac{k_i k_j}{k^2} \delta(kc(1-\frac{1}{2\gamma^2})[\sin{\theta'}\sin{\theta}\cos{\varphi'}+\cos{\theta'}\cos{\theta}]-\omega_p).
\end{equation}
After transforming the delta function we get 
\begin{equation}
    D_{ij}=\pi e^2 \int k^2 dk \int d\cos{\theta'} \int d\varphi' W(\boldsymbol{k},t) \frac{k_i k_j}{k^2}
    \frac{\delta(\varphi'-\varphi'_{*})}{kc(1-\frac{1}{2\gamma^2})\sin{\theta'}\sin{\theta}\sin{\varphi'_*}},
\end{equation}
where $\cos{\varphi'_*} = \frac{\omega_p/(kc(1-\frac{1}{2\gamma^2}))-\cos{\theta'}\cos{\theta}}{\sin{\theta'}\sin{\theta}}$.

$k_i$ is the projection of wave-vector ($\boldsymbol{k} = k\sin{\theta'}\cos{\varphi'}\boldsymbol{\hat{x}}+k\sin{\theta'}\sin{\varphi'}\boldsymbol{\hat{y}}+k\cos{\theta'}\boldsymbol{\hat{z}}$) to the spatial direction $i$. We have fixed the azimuth angle of the pair beam to zero $(\varphi=0)$, therefore we have only the beam modulus momentum ($p$) and the angler direction $\boldsymbol{\hat{\theta}}=\cos{\theta} \boldsymbol{\hat{x}}-\sin{\theta} \boldsymbol{\hat{z}}$. Based on this we find that $k_p$ is the modulus of the wave-vector  and $k_\theta = \boldsymbol{k}\cdot\boldsymbol{\hat{\theta}} = k[\sin{\theta'}\cos{\theta}\cos{\varphi'}-\cos{\theta'}\sin{\theta}]$.

Substituting the values of $k_p$ and $k_\theta$ and integrating over $\varphi'$ gives
\begin{equation}\label{eq:Ma}
            \left\{\begin{array}{lr}
        D_{pp} \\
        D_{p\theta} \\
        D_{\theta \theta} 
        \end{array}\right\} =  \pi \frac{m_e \omega_p^2}{n_e} \int_{\omega_p/c}^\infty k^2dk \int_{\cos{\theta_1'}}^{\cos{\theta_2'}} d\cos{\theta'} \frac{W(\boldsymbol{k}) }{kc(1-\frac{1}{2\gamma^2})\sqrt{(\cos{\theta'}-\cos{\theta'_1})(\cos{\theta'_2}-\cos{\theta'})}} 
        \left\{\begin{array}{lr}
        1 \\
        \xi \\
        \xi^2 
        \end{array}\right\},
\end{equation}
where
\begin{equation}
    \xi  = \sin{\theta'}\cos{\varphi'_*}\cos{\theta} - \cos{\theta'}\sin{\theta} = \frac{\cos{\theta}\frac{\omega_p}{kc(1-\frac{1}{2\gamma^2})}-\cos{\theta'}}{\sin{\theta}}. 
\end{equation}
and the boundaries of $\cos{\theta'}$ are fixed by the condition
\begin{equation}
    \abs{\cos{\varphi'_*}} = \abs{ \frac{\omega_p/(kc(1-\frac{1}{2\gamma^2}))-\cos{\theta'}\cos{\theta}}{\sin{\theta'}\sin{\theta}}} \leq 1,
\end{equation}
which gives 
\begin{equation}\label{eq:cost12}
    \cos{\theta_{1,2}'} = \frac{\omega_p}{kc(1-\frac{1}{2\gamma^2})}\left[\cos{\theta}\pm\sin{\theta}\sqrt{\left(\frac{kc}{\omega_p}\right)^2 (1-\frac{1}{2\gamma^2})^2-1}\right]. 
\end{equation}

Since we have the calculations for the linear growth rate in the Cartesian coordinates $(k_\perp,\epsilon_{||})$ where $k_{||} = \frac{\omega_p}{c}(1+\epsilon_{||})$, we need to transform the diffusion coefficients integrand from the polar $(k,\cos{\theta'})$ to the Cartesian $(k_\perp,\epsilon_{||})$. Multiplying eq.(\ref{eq:Ma}) by the Jacobian determinant $|J| = \frac{\omega_p}{c}\frac{k_\perp}{k^2}$, we get 

\begin{equation}\label{eq:skperp}
            \left\{\begin{array}{lr}
        D_{pp} \\
        D_{p\theta} \\
        D_{\theta \theta} 
        \end{array}\right\} =  \pi \frac{m_e \omega_p^2}{n_e c} \int d\epsilon_{||} \int_{k_{\perp,1}}^{\infty} dk_\perp k_\perp
        \frac{W(k_\perp,\epsilon_{||}) }{\sqrt{\theta^2\left(\frac{k_{\perp}}{\omega_p/c}\right)^2+\epsilon_{||}\left[\theta^2+\frac{1}{\gamma^2}\right]-\epsilon_{||}^2 -\left[\frac{1}{2\gamma^2}+\frac{\theta^2}{2}\right]^2}} 
        \left\{\begin{array}{lr}
        1 \\
        \xi \\
        \xi^2 
        \end{array}\right\},
\end{equation}
where for $\theta<<1$ and $\epsilon_{||}<<1$, we can approximate $\xi$ as
\begin{equation}
    \xi = -\frac{\omega_p}{kc}\frac{1}{\theta}\left[\frac{\theta^2}{2}+\epsilon_{||}-\frac{1}{2\gamma^2}\right].
\end{equation}

The resonance boundaries translate to a lower bound on $k_\perp$ for a given $\epsilon_{||}$. The modes with negative $\epsilon_{||}$ are stable and therefore we are only left with the lower limit for the positive $\epsilon_{||}$ that is given by
\begin{equation}\label{eq:limitkperp}
    k_{\perp,1} = \frac{\omega_p}{c}\frac{1}{\theta}\sqrt{\epsilon_{||}^2 +\frac{1}{4\gamma^4} +\frac{\theta^4}{4} -\frac{\epsilon_{||}}{\gamma^2} +\frac{1}{2}\left(\frac{\theta}{\gamma}\right)^2 - \epsilon_{||}\theta^2}.
\end{equation}

In order to have a proper numerical girding over the unstable waves spectrum we transform the coordinates from $(k_\perp,\epsilon_{||})$ to $(k_\perp,\theta^{R})$ where $\theta^{R} = \frac{\epsilon_{||}}{ck_\perp/\omega_p}$ finding the following final expression for the diffusion coefficients

\begin{equation}\label{eq:DFinal}
            \left\{\begin{array}{lr}
        D_{pp} \\
        D_{p\theta} \\
        D_{\theta \theta} 
        \end{array}\right\} =  \pi \frac{m_e \omega_p^2}{n_e c \theta} \int_{R(\theta,\gamma)} dk_\perp k_{\perp} \int_{R(\theta,\gamma)} d\theta^{R}
        \frac{W(k_\perp,\theta^{R})}{\sqrt{1 -\left(\frac{\theta^R}{\theta}\right)^2 +\frac{\theta^R}{ck_{\perp}/\omega_p}\left[1 +\left(\frac{1}{\gamma\theta}\right)^2\right]- (\frac{\omega_p}{ck_{\perp}})^2\left[\frac{1}{2\gamma^2\theta}+\frac{\theta}{2}\right]^2}} 
        \left\{\begin{array}{lr}
        1 \\
        \xi \\
        \xi^2 
        \end{array}\right\},
\end{equation}

where
\begin{equation}
    \xi =  -\frac{1}{\sqrt{1+2\theta^R (ck_\perp/\omega_p)+(ck_\perp/\omega_p)^2(1+{\theta^{R}}^2)}}\left[\frac{\theta^R}{\theta}\frac{ck_\perp}{\omega_p}+\frac{\theta}{2}-\frac{1}{2\theta\gamma^2}\right],
\end{equation}
and the resonance region $R(\theta,\gamma)$ is defined by the following condition
\begin{equation}
    \left(\frac{ck_{\perp}}{\omega_p}\right)^2 \left(\theta^2-{\theta^{R}}^{2}\right) +\frac{ck_{\perp}}{\omega_p} \theta^R \left[\theta^2+\frac{1}{\gamma^2}\right]-\left[\frac{1}{2\gamma^2}+\frac{\theta^2}{2}\right]^2\ge 0.
\end{equation}

 \section{The pair beam momentum distribution function}\label{app:fbg}

\begin{figure}
\centering
        \includegraphics[width=0.8\columnwidth]{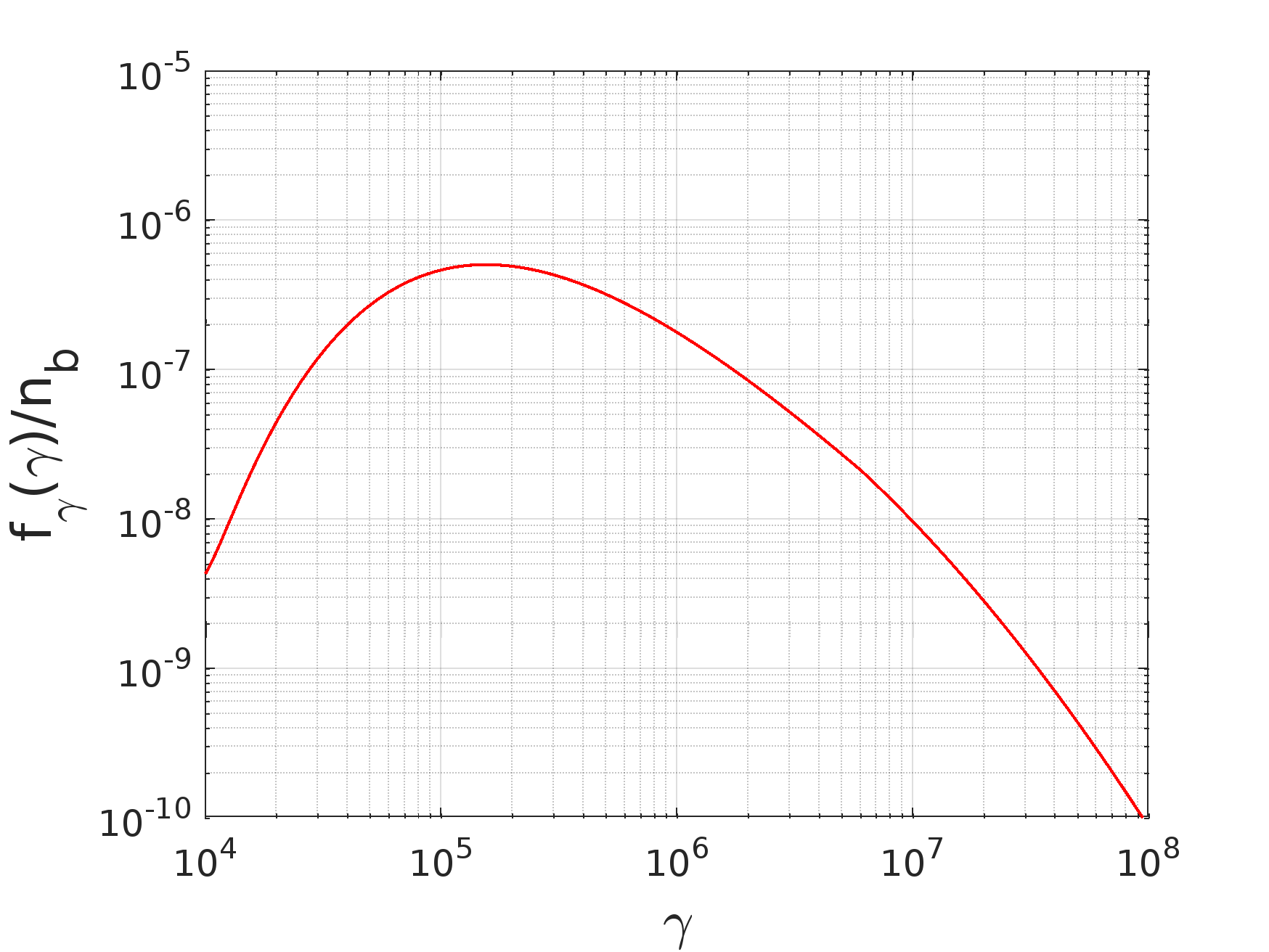}
        \caption{The momentum distribution of the pair beam that we have used in this study as its given by eq.\ref{eq:fbg}.}
        \label{fig:fbg}
\end{figure}

We approximated the pair beam momentum distribution function found in \cite{Vafin_2018} with a logarithmic Gaussian at Lorentz factors higher than $6\times10^6$. This replaces the step-function cut-off with an exponential one. This additional function has continuity in derivative and value at the transition point with the distribution found in \cite{Vafin_2018}, where the resulting pair beam distribution function is given by

\begin{equation}\label{eq:fbg}
    f_\gamma(\gamma) = N_1 \left(\frac{\gamma}{\gamma_1}\right)^{-b1} \exp{-\sqrt{\frac{\gamma_1}{\gamma}}} \Theta\left[(\gamma-6\times 10^3)(6\times10^6-\gamma)\right] + N_2 \left(\frac{\gamma}{\gamma_2}\right)^{-\frac{\ln{(\gamma/\gamma_2)}}{b_2}-1} \Theta\left[(\gamma-6\times 10^6)(10^8-\gamma)\right],
\end{equation}
where the parameters are summarized in Table \ref{tab:fbg}. We have plotted the pair beam distribution function in Fig.\ref{fig:fbg}.

% \usepackage{graphicx}
\begin{table}[]
\centering
\caption{The parameters for the approximation in eq.\ref{eq:fbg}.}
\label{tab:fbg}
\resizebox{0.8\columnwidth}{!}{%
\begin{tabular}{|c|c|c|c|}
\hline
$i$ & $b_i$ & $\gamma_i$         & $N_i$               \\ \hline
1   & 1.60  & $1.58\times10^{6}$ & $3.00 \times 10^{-7}$  \\ \hline
2   & 5.78  & $1.55\times10^{6}$ & $1.14\times10^{-7}$ \\ \hline
\end{tabular}%
}
\end{table}

\bibliographystyle{aasjournal}
\bibliography{awashra_refs} 

\end{document}